\documentclass[11pt, leqno]{article}
\usepackage[paperwidth=8.5in,paperheight=11in,top=1in, bottom=1in, left=0.75in, right=0.75in]{geometry}

\usepackage{amsfonts,amsthm,amssymb,amsmath}
\allowdisplaybreaks
\usepackage{xcolor}
\usepackage{tikz}
\usepackage{bm}
\usepackage{epstopdf}
\usepackage{graphicx}
\usepackage{array}
\usepackage{mathtools}
\mathtoolsset{showonlyrefs=true}

\usepackage{multirow}
\usepackage[authoryear]{natbib}

\usepackage{hyperref} 
\hypersetup{
	colorlinks   = true,
	citecolor    = blue,
	linkcolor    = red,
	urlcolor     = magenta
}

\newtheorem{theorem}{Theorem}[section]

\newtheorem{proposition}[theorem]{Proposition}
\newtheorem{remark}[theorem]{Remark}

\newtheorem{problem}[theorem]{Problem}
\newtheorem{definition}[theorem]{Definition}
\newtheorem{corollary}[theorem]{Corollary}

\numberwithin{equation}{section}

\newcommand{\Eb}{\mathbb{E}}
\newcommand{\Fb}{\mathbb{F}}
\newcommand{\Pb}{\mathbb{P}}
\newcommand{\Rb}{\mathbb{R}}
\newcommand{\Vb}{\mathbb{V}}

\newcommand{\Ac}{\mathcal{A}}
\newcommand{\Fc}{\mathcal{F}}
\newcommand{\Jc}{\mathcal{J}}
\newcommand{\Lc}{\mathcal{L}}

\newcommand{\Vc}{\mathcal{V}}
\newcommand{\Wc}{\mathcal{W}}

\newcommand{\Xt}{X^u_{t,x}}
\newcommand{\Yt}{Y^u_{t,y}}
\newcommand{\bmu}{\bar{\mu}}
\newcommand{\bp}{\bar{p}}

\newcommand\dd{\mathrm{d}}

\usepackage{bbm}

\newcommand{\wh}{\widehat}
\newcommand{\wt}{\widetilde}
\newcommand{\mb}{\mathbf{m}}

\newcommand{\pre}{\mathrm{pre}}

\begin{document}

\title{\Large \scshape Mean-Variance Investment and Risk Control Strategies\\ --
	A Time-Consistent Approach via A Forward Auxiliary Process}

\author{
Yang Shen\thanks{School of Risk and Actuarial Studies, University of New South Wales, Sydney, NSW 2052, Australia. Email: y.shen@unsw.edu.au.}
\and
Bin Zou\thanks{Corresponding author.   Department of Mathematics, University of Connecticut, 341 Mansfield Road U1009, Storrs 06269-1009, USA. Email: bin.zou@uconn.edu. Phone: +1-860-486-3921.}
}

\date{First Version: July 5, 2020; 
This Version: November 29, 2020\\
Accepted for publication in \emph{Insurance: Mathematics and Economics}}
\maketitle

\vspace{-1cm}
\begin{abstract}
\noindent
We consider an optimal investment and risk control problem for an insurer under the mean-variance (MV) criterion. By introducing a deterministic auxiliary process defined forward in time, we formulate an alternative time-consistent problem related to the original MV problem, and obtain the optimal strategy and the value function to the new problem in closed-form.
We compare our formulation and optimal strategy to those under the precommitment and game-theoretic framework.
Numerical studies show that, when the financial market is negatively correlated with the risk process, optimal investment may involve short selling the risky asset and, if that happens, a less risk averse insurer short sells more risky asset.
\end{abstract}



\section{Introduction}
\label{sec:intro}

Investing premiums in financial assets and managing risk from underwriting are two fundamental business decisions to an insurer.
The close interaction between these two decisions motivates us to model a combined financial and insurance market for an insurer and study them simultaneously.
With this in mind, we set up such a combined market consisting of one risk-free asset, one risky asset,
and one risk process $R$ representing the liabilities per unit (or per policy).
We apply a jump-diffusion model for the risk process $R$, which is  a generalization
of the diffusion approximation model and the classical Cram\'er-Lundberg (CL) model.
Such a modeling framework is general in the sense that it covers risk process models used in related literature
as special cases; see, e.g., \cite{browne1995optimal} and \cite{hojgaard1998optimal} for early works without jumps and \cite{zeng2013time} and \cite{zeng2016robust} for more recent works with jumps.
The setup of the combined market in this paper mainly follows from that in \cite{zou2014optimal}; see further motivations in \cite{stein2012stochastic}[Chapter 6] and recent papers \cite{peng2016optimal} and \cite{bo2017optimal}.
We assume the insurer can directly control her liability exposure, or alternatively, the insurer can  decide the total amount of liabilities, measured by units (or the number of policies) $L$ times $R$. One can easily see that such an assumption is equivalent to allowing the insurer to purchase proportional reinsurance to manage her risk exposure from underwriting.
Please refer to \cite{hojgaard1998optimal} and \cite{schmidli2001optimal} for an excellent introduction on optimal proportional reinsurance.
The insurer in consideration then decides the amount $\pi$ to be invested in the risky asset and the liability units $L$ in the insurance business, with the purpose to achieve her optimization objective.

In the control literature within actuarial science,
there are three popular choices for the optimization objective: (1) utility maximization, cf. \cite{yang2005optimal}, \cite{bai2008optimal}, and \cite{liang2011optimal};
(2) risk minimization (e.g., ruin probability, VaR, and CVaR), cf. \cite{schmidli2001optimal}, \cite{liu2004optimal}, and \cite{david2005minimizing};  and (3) the mean-variance (MV) criterion, with the last one used in this paper.
Please see a recent review article \cite{cai2020optimal}, the monograph \cite{albrecher2017reinsurance}, and the references therein for the rich literature on optimal reinsurance under the first two objectives.
MV portfolio selection (without risk control or reinsurance) is first studied by \cite{markowitz1952portfolio} and is truly a cornerstone of the modern portfolio theory.
There are so many works extending \cite{markowitz1952portfolio} along numerous directions that it is almost impossible to give credits to all of them even in a review article.
Here we only mention \cite{li2000optimal}, \cite{zhou2000continuous},
\cite{basak2010dynamic}, \cite{bjork2010general}, and \cite{bjork2014mean} for a short list.
It is well known that a standard dynamic MV problem is a time-inconsistent control problem, in the sense that an optimal strategy obtained at time $t$ may cease to be optimal at a later time $t_1$.
In order for such a strategy to be followed after the initial time, the agent needs to commit herself to it.
Hence, this type of optimal strategy is called the \emph{precommitment} strategy in the literature; see \cite{zhou2000continuous}.
However,
a rational agent who is aware of such a time-inconsistency issue should search for a time-consistent \emph{equilibrium} strategy, which she has no incentive to deviate from once obtained; see \cite{basak2010dynamic} and \cite{kryger2010some}.
For a complete analysis on a more general framework in both discrete and continuous time models, we refer readers to the influential work of \cite{bjork2010general}.
In the coming paragraph, we shall focus on optimal MV (proportional) reinsurance problems and provide a selective literature review on the topic.

\cite{bauerle2005benchmark} is among the early contributions to optimal MV reinsurance problems.
The author applies the embedding technique and the standard Hamilton-Jacobi-Bellman (HJB) approach to obtain the  precommitment proportional reinsurance strategy
under the CL model.
\cite{bai2008dynamic} further add a no short-selling constraint on the investment strategy and solve for the precommitment strategy under both the CL model and the diffusion model for the risk process.
\cite{shen2014optimal} introduce delay into the controlled portfolio and apply a stochastic maximal principle to handle such a problem.
\cite{shen2015optimal} propose an asset model in which
both the appreciation and volatility of the risky asset follow non-Markovian processes, and apply
 a backward stochastic differential equation (BSDE) approach to obtain the precommitment strategy.
On the other hand, time-consistent MV investment-reinsurance problems are first studied by \cite{zeng2011optimal} under a standard Black-Scholes type financial market and a diffusion risk process.  They apply the method from \cite{bjork2010general} and obtain the equilibrium strategy by solving an extend HJB system. Zeng and his collaborators further include jumps in both risky assets and the risk process in \cite{zeng2013time} and ambiguity aversion in \cite{zeng2016robust}.
\cite{li2013optimal} allow the risk aversion to be state dependent, similar to that in \cite{bjork2014mean}.
See \cite{li2015time} for extension with stochastic interest rate and inflation risk.
A latest paper by \cite{cao2020optimal} studies the problem in a contagious model where the risk process is given by a self-exciting Hawkes process.
The above mentioned papers consider Markovian, also called feedback or closed-loop, controls in the analysis.
Two recent works \cite{wang2019time} and \cite{yan2020open} consider open-loop controls under  a model with random parameters and a stochastic volatility model, respectively.

This paper lies in the category of time-consistent strategies to MV investment and proportional reinsurance problems.
We summarize the main results and contributions of this paper as follows.
\begin{itemize}
\item We propose an alternative time-consistent formulation to the insurer's original time-inconsistent MV problem, which is different from both the general approach of \cite{bjork2010general} and the special approach of \cite{basak2010dynamic}.
To be precise, we introduce an auxiliary \emph{deterministic} process $Y$ defined \emph{forward} in time and use $Y$ to replace the conditional expectation of the insurer's wealth $X$ in the original objective $J$, leading to a modified MV objective $\Jc$ (see \eqref{eq:obj2} for details).
We then consider an alternative MV problem in which the  \emph{pair} $(X,Y)$ are taken as the state processes and $\Jc$ is the optimization objective.
Because of the introduction of $Y$ and the enlargement of the state space in the definition of $\Jc$ (which together ``kill'' the troubling variance term in $J$), the alternative MV problem is a time-consistent control problem and can be solved by the standard HJB method.
In comparison,  \cite{bjork2010general} introduce an auxiliary \emph{stochastic} process  defined \emph{backward} in time to handle the square term in the MV problems; while \cite{basak2010dynamic} take a clever use of the total variance formula to derive a heuristic HJB equation.
In terms of applicability to time-inconsistent problems, the approach of \cite{bjork2010general} is the most general one and the approach of \cite{basak2010dynamic} is likely the most restrictive one, as it only applies to the MV problems. The approach of this paper is on the special side as \cite{basak2010dynamic}, but it can also handle general non-linear term(s) involving conditional expectation in the objective, other than the square term in the MV objective.

\item The main approach used in this paper follows from \cite{yang2020bellman},  which considers a standard MV portfolio selection problem without risk control in a Black-Scholes market model.
	On the technical level, we extend the work of \cite{yang2020bellman} by including an additional risk control strategy for an insurer whose risk process is modeled by a jump-diffusion process and is correlated with the risky asset in the financial market.
	Such an extension leads to many interesting findings along both analytical and numerical directions.

\item We fully solve both the time-inconsistent and the time-consistent MV investment and risk control problems in the paper, i.e., we obtain both the precommitment and the (time-consistent) optimal strategies, 
and the efficient frontier and the value function of both problems in closed-form.
Utilizing these explicit results, we conduct a comprehensive analysis to compare our optimal and precommitment strategies with those obtained in the standard game-theoretic framework (see \cite{bjork2010general}) and the precommitted framework (see \cite{zhou2000continuous}).

\item We also make contributions to the literature in the direction of economic analysis on the insurer's optimal strategies, which is missing or insufficient in many related works including \cite{yang2020bellman}. We discuss how the model parameters and the insurer's risk profiles affect the optimal strategies both analytically (see Section \ref{sub:econ}) and numerically (see Section \ref{sec:nume}). In particular, we find that the correlation between the financial market and the insurance market (risk process) plays a key role in the insurer's optimal decisions.
Both analytical results (see Table \ref{tab:impact})  and numerical findings (see Figure \ref{fig:rho_op}) show that when the correlation coefficient $\rho$ is positive, the insurer increases her investment in the risky asset and holds more liabilities from underwriting as $\rho$ increases.
When $\rho$ is negative, we obtain several interesting findings. First, the optimal liability strategy is no longer monotone with respect to $\rho$, but rather has a convex relation, decreasing first and then increasing. Next, when $\rho$ is very negative (close to -1), the optimal investment involves short selling the risky asset and a less risk averse insurer short sells more risky asset.
\end{itemize}

We organize the rest of this paper as follows.
In Section \ref{sec:prob}, we present the market model and state the insurer's MV investment and risk control problems.
In Section \ref{sec:main}, we obtain explicit solutions to the insurer's MV problems.
We then compare our formulation, approach, and optimal strategy with those under the game-theoretic and precommitted framework in Section \ref{sec:comp}.
Section \ref{sec:nume} contains our numerical studies which focus on the impact of correlation and jumps on the optimal strategy.
Our conclusions are summarized in Section \ref{sec:conl}.
Appendix \ref{app:pre} collects technical proofs.

\section{The Framework}
\label{sec:prob}

\subsection{The Market Model}

Let us fix a complete filtered probability space $(\Omega, \Fc, \Fb=(\Fc_t)_{t \in [0,  T]}, \Pb)$ over a finite time horizon $[0,T]$, with $T < \infty$.
Here $\Fc_t$ contains all the information up to time $t$ and the filtration $\Fb$ satisfies the usual hypotheses. We interpret $\Pb$ as the physical probability measure. We assume all the stochastic processes below are well defined and adapted to the given filtration $\Fb$.

We consider an insurer who makes business decisions in a  combined financial and insurance market, similar to the one introduced in \cite{zeng2011optimal} and \cite{zou2014optimal}.
In the financial market, there is one risk-free asset and one risky asset (a stock or an index) whose price dynamics are given respectively by
\begin{align}
\dd S_0(t) &= r S_0(t) \, \dd t,  & S_0(0) &=1, \label{eq:dS0}\\
\dd S_1(t) &= S_1(t) \big(\mu \, \dd t + \sigma \, \dd W_1(t) \big), &  S_1(0) &> 0,  \label{eq:dS1}
\end{align}
where $r,\mu,\sigma>0$ are positive constants and $W_1$ is a standard Brownian motion.
Here, $r$ is the risk-free interest rate, and $\mu$ and $\sigma$ are the appreciation rate and volatility of the risky asset, respectively.
The insurer can dynamically trade both assets without frictions and taxes.
In the insurance market, we model the insurer's unit liabilities (risk) $R=(R(t))_{t \in [0,T]}$ by the following jump-diffusion process:
\begin{align}
\label{eq:dR}
\dd R(t) = \alpha \, \dd t + \beta \left(\rho \, \dd W_1(t) + \sqrt{1 - \rho^2} \, \dd W_2(t) \right) + \int_{\Rb} \, \gamma(t,z) \, N(\dd t, \dd z) ,
\end{align}
where $\alpha, \beta \ge 0$ and $\rho \in [-1,1]$, $W_2$ is another standard Brownian motion independent of $W_1$, $N$ is the Poisson random measure, and $\gamma(t,z) >0$ for all $t$ and $z$.
Here, $\rho$  captures the correlation between the financial market and the insurance market.

We impose several technical assumptions on the models \eqref{eq:dS0}-\eqref{eq:dR} that will be enforced throughout the rest of the paper.
Suppose the compensated Poisson random measure $\wt N$ is given by
\begin{align}
\label{eq:dNt}
 \wt N(\dd t, \dd z) = N(\dd t, \dd z) - \lambda \,  \dd t \, \dd F_Z(z), 
\end{align}
where $\lambda>0$ denotes the jump intensity and $F_Z$ is the distribution function of a random variable $Z$. In addition, $\gamma(t,\cdot) = \gamma(\cdot)$ is homogeneous and deterministic, and $\gamma(Z)$ has finite first and second moments, i.e.,
\begin{align}
\label{eq:gam-moments}
\overline \gamma_1 := \int_{\Rb} \, \gamma(z) \, \dd F_Z(z) \in (0, \infty) \qquad \text{and} \qquad
\overline{\gamma}_2 := \int_{\Rb} \,  \gamma^2(z)  \, \dd F_Z(z) \in (0, \infty).
\end{align}
We assume $W_1$, $W_2$, $N$, and $Z$ are stochastically independent, and the filtration $\Fb$ is generated by them and augmented with $\Pb$-null sets.

In the financial market, the insurer chooses an investment strategy $\pi = (\pi(t))_{t \in [0,T]}$, where $\pi (t)$ denotes the amount of wealth invested in the risky asset at time $t$.
In the insurance market, the insurer chooses a risk control strategy (or a liability strategy) $L = (L(t))_{t \in [0,T]}$, where $L (t)$ denotes the amount of liabilities in the underwriting at time $t$.
Assume the unit premium rate, corresponding to the unit liabilities (risk) $R$, is given by $p$, where $p>0$.
For any fixed risk control strategy $L$, the gains from the insurance business evolve according to $L(t) \big(p \, \dd t - \dd R(t) \big)$.

\begin{remark}
	\label{rem:model}
In the combined market model \eqref{eq:dS0}-\eqref{eq:dR}, we set the model parameters ($r$, $\mu$, $\sigma$,  $\alpha$, $\beta$, and $\lambda$) to be positive constants.
We comment that all the analysis and results in the sequel hold if these parameters are given by deterministic, bounded, and positive processes, and the volatility process $\sigma$  is bounded away from zero (i.e., there exists a positive constant $K$ such that $\sigma(t) \ge K >0$).
Similarly, if the deterministic function $\gamma(t,\cdot)$ is not time homogeneous, we need to replace $\overline{\gamma}_i$ in \eqref{eq:gam-moments} by $\overline{\gamma}_i(t)$, and assume $\overline{\gamma}_i(t)$ are bounded for all $t$, where $i=1,2$.
Given that all the parameters are constants and the function $\gamma$ is homogeneous, we set the unit premium rate to be a positive constant $p$.
The above assumptions on modeling, albeit strong, are standard and popular in the related literature; see, e.g, \cite{schmidli2001optimal}, \cite{yang2005optimal}, \cite{moore2006optimal}, and \cite{zeng2011optimal}.

The risk model \eqref{eq:dR} incorporates several well known models in actuarial science. If we set $\alpha = \beta =0$, then the model \eqref{eq:dR} is a generalization of the classical Cram\'er-Lundberg (CL) model.
Recall the risk process $R(t)$ in the CL model is given by $R(t) = \sum_{i=1}^{\wh{N}(t)} \, \wh{C}_i$, where $\wh{N} = (\wh{N}(t))_{t \in [0,T]}$ is a homogeneous Poisson process with constant intensity $\wh{\lambda}$ and $(\wh{C}_i)_{i=1,2,\cdots}$ is a series of independent and identically distributed random variables, also independent of $\wh{N}$. Comparing \eqref{eq:dR} with the CL model shows that (i) $\lambda = \wh \lambda$ and (ii) $\gamma(t,Z)$ and $\wh C$ has the same distribution.
If we set $\lambda=0$, i.e., no jumps in \eqref{eq:dR}, then the model \eqref{eq:dR} can be seen as a diffusion approximation to the CL model; see, e.g.,  \cite{browne1995optimal}, \cite{hojgaard1998optimal}, and \cite{moore2006optimal}. In such a case, we have  $\alpha = \wh{\lambda} \, \Eb[\wh{C}]$ and $\beta^2 = \wh{\lambda} \, \Eb[\wh{C}^2]$.

In our framework, we interpret $L$ as the amount of liabilities the insurer decides to take in the insurance business, and $p$ as the premium rate the insurer receives from underwriting the policies against the risk $R$.
This modeling choice follows from \cite{stein2012stochastic}[Chapter 6] and its subsequent studies such as \cite{zou2014optimal}, \cite{peng2016optimal}, and \cite{bo2017optimal},
where the motivation comes from the AIG case in the financial crisis of 2007-2008 and argues for a negative correlation $\rho < 0$.

It is clear that our risk control setup is consistent with the model of proportional reinsurance, an important topic in actuarial science.
In the latter case, we should understand $R$ in \eqref{eq:dR} as the risk process of the insurer, and $p$ the reinsurance premium paid by the insurer to the reinsurer.
To manage risk, the insurer chooses proportional insurance, with $L$ denoting the retention proportion.
The dynamics of the gains process are then given by $L(t) \, \dd R(t) - p (1 - L(t)) \, \dd t$.
Indeed, our setup (with $\lambda = 0$) is the same to that in \cite{zeng2011optimal} by taking $m=0$ and $\sigma_0$ to be the negative value there.
\end{remark}

In the combined market described above, let us introduce $u = (\pi, L)$ as a shorthanded notation for the insurer's control or strategy.
As usual, we consider self-financing strategies only in the analysis.
For a fixed strategy $u$, we write $X^u = (X^u(t))_{t \in [0,T]}$ as the insurer's wealth process, and obtain
the dynamics of $X^u$  by
\begin{align}
\label{eq:dX}
\dd X^u(t) = & \big(rX^u(t)  + \bmu \pi(t) + \bp  L(t) \big) \, \dd t  + \big(\sigma \pi(t) - \rho \beta L(t) \big) \, \dd W_1(t) \\
&- \beta \sqrt{1 - \rho^2} L(t) \, \dd W_2 -  L(t) \, \int_{\Rb} \, \gamma(z) \, N(\dd t, \dd z),
\end{align}
where the initial wealth $X(0)$ is  a (positive) constant, and $\bmu$ and $\bp$ are defined by
\begin{align}
\label{eq:bar}
\bmu:= \mu - r \qquad \text{ and } \qquad \bp := p - \alpha.
\end{align}
If we fix an initial state $X^u(t) = x$ at time $t$, where $t \in [0,T]$,  and want to emphasize the dependence of $X^u$ on the initial state, we use the notation $(X^u_{t,x}(s))_{s \in [t,T]}$.

To make sure the stochastic differential equation (SDE) \eqref{eq:dX} admits a unique strong solution, we need to impose square integrability conditions on strategies. Given the nature of the Markovian framework, we consider Markov (feedback) controls, $\pi(t) = \wt{\pi}(t, X^u(t))$ and $L(t) = \wt{L}(t, X^u(t))$, for some deterministic functions $\wt{\pi}$ and $\wt{L}$.
We are now ready to state the admissible set of strategies, denoted by $\Ac$, as follows.

\begin{definition}
	\label{def:adm}
A strategy $u$ is called admissible if (1) $u$ is progressively measurable with respect to the underlying filtration $\Fb$,
(2) $\Eb[\int_0^T \, \pi^2(t) \, \dd t] < \infty$ and  $\Eb[\int_0^T \, L^2(t) \, \dd t] < \infty$, (3) $u$ is a feedback control, and (4)
$X^u$ satisfies the SDE \eqref{eq:dX}.
\end{definition}

\begin{remark}
In defining the admissible set, we do not impose $L \ge 0$. In other words, we allow $L<0$,  which corresponds to the case of taking a strategy greater than 1 in the proportional reinsurance and is often interpreted as acquiring new businesses in the literature; see, e.g., \cite{bauerle2005benchmark} and \cite{zeng2011optimal}.
If we insist on $L \ge 0$, we can impose an extra condition on the model parameters, so that the optimal liability strategy $L^*$ is always non-negative; see  Eq.(6) in \cite{zou2014optimal}.
\end{remark}

\subsection{The Problem}

We consider a representative insurer who is a  mean-variance (MV) type agent. Namely, the insurer prefers higher mean and lower variance of her terminal wealth. Following the standard literature of \cite{li2000optimal} and \cite{zhou2000continuous}, we define the insurer's objective functional $J$ by
\begin{align}
\label{eq:obj}
J(t,x; u) := \Eb_{t,x} \left[X^u(T)\right] - \frac{\theta}{2} \, \Vb_{t,x} \left[X^u(T)\right],
\end{align}
where $\theta > 0$ is the risk aversion parameter, $X^u$ is given by \eqref{eq:dX}, and $\Eb_{t,x}$ (resp. $\Vb_{t,x}$) denotes taking conditional expectation (resp. variance) given $X^u(t) = x$ under the physical measure $\Pb$. We now state the insurer's MV investment and risk control problem as follows.

\begin{problem}[A Time-Inconsistent MV Problem]
	\label{prob:incons}
The insurer seeks a strategy $u^\pre=(u^\pre(s))_{s \in [t,T]}$ to maximize the objective functional $J$ defined in \eqref{eq:obj}, i.e., the insurer solves the following MV problem:
\begin{align}
	\label{eq:prob}
	V(t,x) := \sup_{u \in \Ac} \; J(t,x ; u).
\end{align}
We call $V$ the value function to Problem \eqref{eq:prob}.
\end{problem}

As is well known in the literature, a solution to Problem \eqref{eq:prob} is \emph{time-inconsistent}, and hence the insurer has the incentive to deviate from such a strategy at a later time $s > t$. As such, a strategy solving Problem \eqref{eq:prob} will be followed over the remaining period $[t, T]$ only if the insurer commits to it. For this reason, we call a solution to Problem \eqref{eq:prob} a \emph{precommitment} strategy with notation $u^\pre$.
Time-consistent formulations and approaches to the MV problems are then proposed and investigated, while most, if not all,  of them follow the influential work of \cite{bjork2010general}.
However, we will take a different approach from \cite{yang2020bellman} in this paper, which is, although less general, simple and sufficient to handle the problem under our framework.

Inspired by \cite{yang2020bellman}, for any $u = (\pi, L) \in \Ac$, we introduce an auxiliary process
$(\Yt(s))_{s \in [t,T]}$,
which is defined \emph{forward} in time by
\begin{align}
	\label{eq:dY}
	\dd \Yt(s)  = \left( r \Yt(s) + \bmu \, \Eb_{t,y} \big[\pi(s) \big] + (\bp  - \lambda \overline{\gamma}_1) \, \Eb_{t,y} \big[L(s) \big] \right) \dd s,
	\qquad s \in [t, T],
\end{align}
where $\Yt(t) = y $ is the initial state for arbitrary but fixed $t \in [0,T]$ and $y \in \Rb$, $\overline{\gamma}_1$ is given by \eqref{eq:gam-moments}, and $\bmu$ and $\bp$ are defined in \eqref{eq:bar}.
Here, the initial state value $y$ may be different from that of $\Xt(t) = x$.
Using \eqref{eq:dX}, we obtain that
\\[-8ex]
	\begin{align}
		\Eb_{t,y}[X^u_{t,y}(s)] &= y + \Eb_{t,y} \left[\int_t^s \, \left(r X_{t,y}^u(v) + \bmu \pi(v) + (\bp - \lambda \overline{\gamma}_1) L(v) \right) \dd v\right] + \Eb_{t,y} \left[\int_t^s \, \big(\sigma \pi(v) - \beta \rho L(v) \big) \dd W_1(v) \right] \\
		&\quad - \Eb_{t,y} \left[\int_t^s \, \beta \sqrt{1 - \rho^2} L(v) \, \dd W_2(v) \right]
		- \Eb_{t,y} \left[\int_t^s \int_{\Rb} \,  L(v)  \gamma(z)\,   \wt{N}(\dd v, \dd z) \right] \\
		&= y + \int_t^s \, \left(r \, \Eb_{t,y} \left[X_{t,y}^u(v)\right] + \bmu \, \Eb_{t,y} \big[ \pi(v) \big] + (\bp - \lambda \overline{\gamma}_1) \Eb_{t,y} \big[ L(v) \big] \right) \dd v, \qquad \forall \, s \in [t, T],
	\end{align}
	where $X_{t,y}^u(t)=y$,
	and $\wt N$ and $\overline{\gamma}_1$ are defined respectively by \eqref{eq:dNt} and \eqref{eq:gam-moments}.
	Here, we have used the square integrability conditions of $\pi$ and $L$ from Definition \ref{def:adm} and $\overline{\gamma}_2 =\Eb[\gamma^2(Z)] <\infty$ in \eqref{eq:gam-moments} to conclude that the (conditional) expectations of the two It\^o integrals and the integral with respect to $\wt{N}$ are zero.
	By comparing the above result with the dynamics of $Y$ in \eqref{eq:dY},
	we see that $Y^u_{t,y}(s) = \Eb_{t,y}[X^u_{t,y}(s)]$ for all $s \in [t,T]$ and $y \in \Rb$.

We now treat both $X$ and $Y$ as state processes and consider a modified objective functional $\Jc$ defined by
\begin{align}
	\label{eq:obj2}
	\Jc(t,x,y;u) = \Eb_{t,x,y} \left[\Xt(T)  - \frac{\theta}{2} \left(\Xt(T) - \Yt(T) \right)^2 \right] ,
\end{align}
where $\theta>0$ is the risk aversion parameter, $X^u$ and $Y^u$ are given respectively by \eqref{eq:dX} and \eqref{eq:dY}, and $\Eb_{t,x,y}$ denotes taking conditional expectation under $\Xt(t)=x$ and $\Yt(t) = y$.
Notice that there is a fundamental difference between $J$ in \eqref{eq:obj} and $\Jc$ in \eqref{eq:obj2}.
That is, we no longer have the ``troubling" square term $\left(\Eb_{t,x}[X^u(T)]\right)^2$, which causes time-inconsistency in Problem \eqref{eq:prob}.
Based on the new objective functional $\Jc$, we formule a time-consistent version of the original Problem \eqref{eq:prob}.

\begin{problem}[A Time-Consistent MV Problem]
The insurer seeks an optimal strategy $u^*$ to maximize the objective functional $\Jc$ defined in \eqref{eq:obj2}, i.e., the insurer solves the following MV problem:
\begin{align}
	\label{eq:prob2}
	\Vc(t,x,y) = \sup_{u \in \Ac} \; \Jc(t,x,y; u).
\end{align}
We call $\Vc$ the value function to Problem \eqref{eq:prob2}.
\end{problem}

\section{Main Results}
\label{sec:main}

In this section, we first solve the insurer's time-inconsistent problem (Problem \eqref{eq:prob}) to obtain a precommitment strategy in Section \ref{sub:incons} and then solve the insurer's time-consistent problem (Problem \eqref{eq:prob2}) to obtain an optimal strategy  in Section \ref{sub:cons}. We discuss the economic implications based on the results of Problem \eqref{eq:prob2} in Section \ref{sub:econ}.

We impose a standing assumption for the subsequent analysis:
\begin{align}
\label{eq:assu}
\beta^2(1-\rho^2) + \lambda \overline{\gamma}_2  \neq 0,
\end{align}
where $\overline{\gamma}_2$ is given by \eqref{eq:gam-moments}.
The assumption in \eqref{eq:assu} is rather weak and  holds in most conditions.
In fact, it only fails when there are (1) no jumps ($\lambda=0$)  and (2) no diffusion term ($\beta=0$) or perfect correlation ($\rho = \pm 1$). 

\subsection{Explicit Solutions to Problem (\ref{eq:prob})}
\label{sub:incons}

In this subsection, we solve the insurer's time-inconsistent MV problem, as formulated in Problem \eqref{eq:prob}, and obtain explicit solutions in the following theorem.

\begin{theorem}
	\label{thm:pre}
	A precommitment strategy to Problem \eqref{eq:prob}, denoted by $u^\pre = (\pi^\pre(s), L^\pre(s))_{s \in [t,T]}$,  is given by
	\begin{align}
		\label{eq:pre_op2}
		\pi^\pre(s) &= - \kappa_1 \left(X^\pre(s) - x \, e^{r(T-t)} - \frac{1}{\theta} \, e^{\kappa_3(T-t)}\right) \qquad \text{ and } \qquad L^\pre(s) = - \frac{\kappa_2}{\kappa_1} \, \pi^\pre(s),
	\end{align}
	where  $X^\pre$ is the wealth process under the precommitment strategy $u^\pre$ and the initial state $(t,x)$, and the constants $\kappa_i$, $i=1,2,3$, are given by
	\begin{align}
		\kappa_1 := \frac{\bmu  (\beta^2 + \lambda \overline{\gamma}_2 ) + \rho \beta \sigma ( \bp - \lambda \overline{\gamma}_1)}{\left( \beta^2(1-\rho^2) + \lambda \overline{\gamma}_2 \right) \sigma^2}, \qquad  \qquad
		&\kappa_2 := \frac{\rho \beta  \bmu + (\bp - \lambda \overline{\gamma}_1 ) \sigma}{\left( \beta^2(1-\rho^2) + \lambda \overline{\gamma}_2 \right) \sigma}, \label{eq:kappa12}
	\end{align}
\begin{align}
	\kappa_3 := \frac{\left(\beta^2 + \lambda \overline{\gamma}_2 \right) \bmu^2 + 2 \rho \beta \sigma \bmu (\bp - \lambda \overline{\gamma}_1) + \left(\bp - \lambda \overline{\gamma}_1 \right)^2 \sigma^2}{\left( \beta^2(1-\rho^2) + \lambda \overline{\gamma}_2 \right) \sigma^2} =
	\frac{\bmu^2}{\sigma^2} + \frac{\left(\bp - \lambda \overline{\gamma}_1 - \frac{\rho \beta \bmu}{\sigma}  \right)^2}{\beta^2(1-\rho^2) + \lambda \overline{\gamma}_2},
	\label{eq:kappa3}
\end{align}
with $\overline{\gamma}_1$ and $\overline{\gamma}_2$ defined in \eqref{eq:gam-moments}, and $\bmu$ and $\bp$ defined in \eqref{eq:bar}.
\end{theorem}	

\begin{proof}
Please refer to Appendix \ref{app:pre} for a proof.
\end{proof}

\begin{proposition}
\label{prop:in-effi}
Let $X^\pre$ be the insurer's wealth process under the precommitment strategy $u^\pre$ given by \eqref{eq:pre_op2}.
We have
	\begin{align}
	\label{eq:pre2_EV}
	\Eb_{t,x} \left[X^\pre(T)\right] = x \, e^{r(T-t)} + \frac{e^{\kappa_3 (T-t) } - 1}{\theta} \qquad \text{ and } \qquad
	\Vb_{t,x} \left[X^\pre(T)\right] = \frac{e^{\kappa_3 (T-t) } - 1}{\theta^2},
\end{align}
where $\kappa_3$ is given in \eqref{eq:kappa3}.
The efficient frontier of Problem \eqref{eq:prob} is obtained by
\begin{align}
	\Vb_{t,x} \left[X^\pre(T)\right] =  \frac{\left( \Eb_{t,x} \left[X^\pre(T)\right] - x \, e^{r(T-t)} \right)^2}{e^{\kappa_3 (T-t) } - 1},
\end{align}
and the value function (mean-variance tradeoff) of Problem \eqref{eq:prob} is given by
\begin{align}
	\label{eq:mv-pre}
	V(t,x) = \Eb_{t, x} \left[ X^\pre(T) \right] - \frac{\theta}{2} \Vb_{t,x} \left[X^\pre(T) \right] = x \, e^{r(T-t)} +  \frac{e^{\kappa_3 (T-t) } - 1}{2\theta} .
\end{align}
\end{proposition}

\begin{proof}
By plugging  $u^\pre$ in \eqref{eq:pre_op2} back into the SDE \eqref{eq:dX}, we obtain the above explicit results.
\end{proof}

\begin{remark}
\label{rem:pre}
From \eqref{eq:pre_op2}, one can easily see that the strategy $u^\pre= (\pi^\pre, L^\pre)$ strongly depends on the initial state $(t,x)$, so a more precise but also more cumbersome notation is to replace it by $u^\pre_{t,x} = (\pi_{t,x}^\pre, L^\pre_{t,x})$.
	Also it is obvious from \eqref{eq:pre_op2} that $u^\pre$ is indeed time-consistent as claimed.
	To see this, let $t < t_1 < t_2 < T$ and $X^\pre_{t,x}(t_1)$ be the corresponding wealth at time $t_1$ under the strategy $u^\pre$. Using \eqref{eq:pre_op2}, we have $u^\pre_{t,x}(t_2) \neq u^\pre_{t_1, X^\pre_{t,x}(t_1)}(t_2)$ in general. That means the ``best" strategy for a future time $t_2$ found at the state $(t,x)$ is \emph{not} the same as the one found at the state $(t_1, X^\pre_{t,x}(t_1))$.
	Here, by the ``best" strategy, we mean a solution to Problem \eqref{eq:prob}.

In Problem \eqref{eq:prob}, $\theta$ is a free parameter, called the insurer's risk aversion parameter, which specifies the insurer's risk attitude towards the mean-variance tradeoff.
Since $\kappa_3 >0$ due to \eqref{eq:kappa3} and $\theta >0$ by definition, we derive from \eqref{eq:pre2_EV} that  there is a one-to-one relation between the target expected terminal wealth  $\mb:=\Eb_{t,x} \left[X^\pre(T)\right]$ and the risk aversion parameter $\theta$, for all $\mb > x \, e^{r(T-t)}$.

\end{remark}

\subsection{Explicit Solutions to Problem (\ref{eq:prob2}) }
\label{sub:cons}

The key to solving Problem \eqref{eq:prob2} is the standard  Hamilton-Jacobi-Bellman (HJB) approach, as presented in Theorem \ref{thm:HJB}.
The proof to this theorem is rather standard in the literature, which is based on the flow property of SDEs, dynamic programming principle, and a verification theorem.
We omit the proof here and refer readers to Theorems 3.3 and 3.4, and Proposition 3.5 in Chapter 4 of  \cite{yong1999stochastic} for a standard proof on a more general control problem.

\begin{theorem}
\label{thm:HJB}
Suppose there exists a classical solution $\Vc$ to Problem \eqref{eq:prob2}. Then $\Vc$ solves the following Hamilton-Jacobi-Bellman (HJB) equation:
\begin{align}
\Vc_t(t,x,y) + \sup_{ \pi, \, L \in \Rb} \;  \bigg\{ (rx + \bmu \pi + \bp L) \Vc_x(t,x,y) + \frac{1}{2} \left[ (\sigma \pi - \rho \beta L)^2 + \beta^2 (1 - \rho^2) L^2 \right] \Vc_{xx}(t,x,y)  \\
+(ry + \bmu \pi + (\bp - \lambda \overline{\gamma}_1 ) L) \Vc_y(t,x,y)  + \lambda \int_{\Rb} \, \big[\Vc(t, x- L \gamma(z), y) - \Vc(t,x,y)\big] \dd F_Z(z)
\bigg\} = 0,  \label{eq:hjb}
\end{align}
for all $(t,x,y) \in [0,T) \times \Rb \times \Rb$, and satisfies the terminal condition
\begin{align}
\label{eq:bound}
\Vc(T,x,y) = x - \frac{\theta}{2} (x-y)^2, \qquad \forall \, x, y \in \Rb.
\end{align}
\end{theorem}

By applying Theorem \ref{thm:HJB}, we obtain explicit solutions to the optimal strategy and the value function  of Problem \eqref{eq:prob2}, as summarized below.

\begin{theorem}
	\label{thm:main}
An optimal strategy to Problem \eqref{eq:prob2}, denoted by $u^*=(\pi^*(s), L^*(s))_{s \in [t,T]}$,  is given by 	
\begin{align}
\label{eq:op}
\pi^*(s) = \frac{\kappa_1}{\theta} e^{-r(T-s)} \qquad \text{ and } \qquad
L^*(s) = \frac{\kappa_2}{\theta}  e^{-r(T-s)}, \qquad s \in [t,T],
\end{align}
where $\kappa_1$ and $\kappa_2$ are defined in \eqref{eq:kappa12}.
The value function $\Vc(t,x,y)$ to Problem \eqref{eq:prob2} is given by
\begin{align}
\label{eq:value}
\Vc(t,x,y) = - \frac{\theta}{2} \, e^{2r(T-t)} (x-y)^2 + e^{r(T-t) } \, x + \frac{\kappa_3}{2 \theta} \, (T-t),
\end{align}
where $\kappa_3$ is defined in \eqref{eq:kappa3}.

\end{theorem}

\begin{proof}
Please see Appendix \ref{app:pre} for a proof.
\end{proof}

\begin{remark}
Since both the optimal investment strategy $\pi^*$ and the optimal liability strategy $L^*$ in \eqref{eq:op} are \emph{independent} of the initial state $(t,x)$, the optimal strategy $u^*$ in \eqref{eq:op} is indeed time-consistent.
We call $u^*$ an optimal strategy, instead of an equilibrium strategy, since the alternative formulated Problem \eqref{eq:prob2} is a standard time-consistent stochastic control problem.

We next comment on possible generalizations to the model \eqref{eq:dS0}-\eqref{eq:dR}.
First, as mentioned in Remark \ref{rem:model}, it is straightforward to extend to the case when all model parameters are deterministic and bounded processes.
Indeed, we simply replace all the constant parameters in $\kappa_i$ by their corresponding (deterministic) process version in $\kappa_i(t)$, where $i=1,2,3$, and the product with time arguments by an appropriate integral, e.g., we change $r (T-t)$ to $\int_t^T \, r(s) \, \dd s$ and $\kappa_3 (T-t)$ to $\int_t^T \, \kappa_3(s) \, \dd s$ in Theorems \ref{thm:pre} and \ref{thm:main}.
Second, we consider only one risky asset in our model, but the analysis and key results in Theorems \ref{thm:pre} and \ref{thm:main} apply to the case of multiple risky assets in a parallel way once we use appropriate matrix notation. In fact, our technique is adequate to handle an incomplete market with $n$ risky assets driven by $d$ independent Brownian motions, where $n \le d$. In such a case, we modify the assumption that $\sigma > 0$ to $\sigma(t) \sigma(t)^\top \ge K \mathbf{I}_{n \times n}$ for some positive $K$ and all $t \in [0,T]$, where $\,^\top$ denotes transpose operator and $\mathbf{I}_{n \times n}$ is an $n\times n$ identity matrix.
\end{remark}

We next derive the efficient frontier of Problem \eqref{eq:prob2} and present the results in the proposition below.

\begin{proposition}
	\label{prop:eff}
	Let $X^*$ be the insurer's wealth process under the optimal strategy $u^*$ given by \eqref{eq:op}.
	We obtain the dynamic efficient frontier of Problem \eqref{eq:prob2} by
	\begin{align}
		\label{eq:eff}
		\Vb_{t,x} \left[X^*(s)\right] = \frac{\left( \Eb_{t,x} \left[X^*(s)\right] - x \, e^{r(s-t)}  \right)^2}{\kappa_3 (s-t)}, \qquad t < s \le T,
	\end{align}
	and the mean-variance tradeoff by
	\begin{align}
		\label{eq:mv-dyma} \quad
		\Eb_{t,x} \left[X^*(s)\right] - \frac{\theta}{2} \Vb_{t,x} \left[X^*(s)\right]  = x \, e^{r(s-t)} + \frac{\kappa_3}{\theta} \, e^{-r(T-s)} \, \left( 1 - \frac{1}{2} e^{-r(T-s)}\right) (s-t)  , \qquad t \le s \le T.
	\end{align}
\end{proposition}

\begin{proof}
	By plugging \eqref{eq:op} into \eqref{eq:dX}, we obtain
	\begin{align}	
		\label{eq:op_W}
		e^{r(T-s)} \, X^*(s) &=  e^{r(T-t)} \, X^*(t) + \frac{\bmu \kappa_1 +  (\bp - \lambda \overline{\gamma}_1) \kappa_2 }{\theta} \,  (s-t)
		+ \frac{\sigma \kappa_1 - \rho \beta \kappa_2}{\theta}  \,  \big(W_1(s) - W_1(t)\big) \\
		&\quad - \frac{\beta \sqrt{1 - \rho^2} \, \kappa_2}{\theta}  \,  \big(W_2(s) - W_2(t)\big) - \frac{\kappa_2}{\theta}    \, \int_t^s \, \int_{\Rb} \, \gamma(z) \, \wt N(\dd v, \dd z),  	
	\end{align}
	where $\kappa_1$ and $\kappa_2$ are defined in \eqref{eq:kappa12}.
	Taking expectation and variance on \eqref{eq:op_W} given the initial state $(t, X^*(t) = x)$, we get
	\begin{align}
		\label{eq:op_EW}
		\Eb_{t,x} \left[X^*(s)\right] &= x \, e^{r(s-t)} + \frac{\bmu \kappa_1 + (\bp - \lambda \overline{\gamma}_1) \kappa_2 }{\theta} \, e^{-r(T-s)} \, (s-t)  \\
		&= x \, e^{r(s-t)} + \frac{\kappa_3 }{\theta} \, e^{-r(T-s)} \, (s-t) , \\
		\label{eq:op_VW}
		\Vb_{t,x} \left[X^*(s)\right] &= \frac{e^{-2r(T-s)} \,  (s-t)  }{\theta^2}   \left[\left(\sigma \kappa_1 - \rho \beta \kappa_2 \right)^2 + \beta^2(1 - \rho^2) \kappa_2^2 + \lambda \overline{\gamma}_2 \kappa_2^2\right]\\
		&= \frac{\kappa_3}{\theta^2} \, e^{-2r(T-s)} \,  (s-t),
	\end{align}
	where we have used the following result
	\begin{align}
		\label{eq:eql}
		\bmu \kappa_1 + (\bp - \lambda \overline{\gamma}_1) \kappa_2 = \left(\sigma \kappa_1 - \rho \beta \kappa_2 \right)^2 + \beta^2(1 - \rho^2) \kappa_2^2 + \lambda \overline{\gamma}_2 \kappa_2^2 = \kappa_3 > 0.
	\end{align}
	We obtain \eqref{eq:eql} by recalling the definitions of $\kappa_i$, $i=1,2,3$, in \eqref{eq:kappa12} and \eqref{eq:kappa3}.
	Combining \eqref{eq:op_EW}, \eqref{eq:op_VW}, and \eqref{eq:eql} leads to the above dynamic efficient frontier of Problem \eqref{eq:prob2}.
\end{proof}

With explicit results obtained in \eqref{eq:op_EW} and \eqref{eq:op_VW}, we can provide a further verification of the optimal investment strategy $u^*$ in \eqref{eq:op}. To this end, let us recall the definition of $Y^u$ in \eqref{eq:dY} and the objective functional $\Jc$ in \eqref{eq:obj2}. Denote $X^*$ and $Y^*$ the corresponding processes under the optimal strategy $u^*$. We obtain
\begin{align}
	\Jc(t,x,y; u^*) &= \Eb_{t,x,y} \left[ X^*_{t,x}(T)  - \frac{\theta}{2} \left(X^*_{t,x}(T) - Y^*_{t,y}(T) \right)^2  \right] \\
	&= \Eb_{t,x} \left[ X^*_{t,x}(T)  \right] -  \frac{\theta}{2} \Vb_{t,x} \left[ X^*_{t,x}(T)  \right] -  \frac{\theta}{2} \left( \Eb_{t,x} \left[ X^*_{t,x}(T)  \right] - \Eb_{t,y} \left[ Y^*_{t,y}(T)  \right] \right)^2 \\
	&= x \, e^{r(T-t)} + \frac{\kappa_3}{\theta} \, (T-t)  - \frac{\kappa_3}{2 \theta} \, (T-t) - \frac{\theta}{2}\left(x \, e^{r(T-t)}  - y \, e^{r(T-t)} \right)^2 \\
	&= \Vc(t,x,y) \quad \text{ derived in \eqref{eq:value}}.
	\label{eq:veri}
\end{align}
Hence, $u^*$ given by \eqref{eq:op} is optimal to Problem \eqref{eq:prob2}.

\subsection{Economic Discussions}
\label{sub:econ}

In this subsection, we present economic discussions on the explicit results of Problem \eqref{eq:prob2} in Theorem \ref{thm:main} and Proposition \ref{prop:eff}.

We first derive two corollaries when the insurer has no access to  the financial market or decides not to take any insurance business. To this purpose, we directly follow the same arguments in the proof of Theorem \ref{thm:main} by setting $\pi \equiv 0$ (or $L \equiv 0$) and obtain the optimal strategy and the value function in each case.
We skip the cumbersome computations and report the results below.

\begin{corollary}
\label{cor:main1}
If the insurer has no access to the financial market ($\pi(s) \equiv 0$ for all $s \in [t,T]$), the optimal risk control strategy $L^*$ to Problem \eqref{eq:prob2} is given by
\begin{align}
L^*(s) \Big|_{\pi \equiv 0} = \frac{\bp - \lambda \overline{\gamma}_1}{(\beta^2 + \lambda \overline{\gamma}_2) \theta} \, e^{-r(T-s)}  , \qquad \forall \, s \in [t,T],
\end{align}
and the value function is given by
\begin{align}
\Vc(t,x,y)\Big|_{\pi \equiv 0}  = - \frac{\theta}{2} e^{2r(T-t)} (x-y)^2 + e^{r(T-t) } x + \frac{\kappa_4}{2 \theta} (T-t), \qquad \text{ where} \quad
\kappa_4 = \frac{(\bp - \lambda \overline{\gamma}_1)^2}{\beta^2 + \lambda \overline{\gamma}_2}.
\end{align}
The loss due to no access to investing in the financial market is measured by
\begin{align}
\Vc(t,x,y) - \Vc(t,x,y)\Big|_{\pi \equiv 0} = \frac{\left( \bmu (\beta^2 + \lambda \overline{\gamma}_2) + \rho \beta \sigma (\bp - \lambda \overline{\gamma}_1)  \right)^2}{(\beta^2 + \lambda \overline{\gamma}_2) (\beta^2(1 - \rho^2) + \lambda \overline{\gamma}_2)} \cdot \frac{T-t}{2 \theta} > 0.
\end{align}
\end{corollary}

\begin{corollary}
\label{cor:main2}
If the insurer sets $L (s)\equiv 0$ for all $s \in [t,T]$,
the optimal investment strategy $\pi^*$ to Problem \eqref{eq:prob2} is given by
\begin{align}
\label{eq:pi_L=0}
\pi^*(s) \Big|_{L \equiv 0} = \frac{\bmu  }{\theta \sigma^2 } \, e^{-r(T-s)}  , \qquad \forall \, s \in [t,T],
\end{align}
and the value function is given by
\begin{align}
\Vc(t,x,y)\Big|_{L \equiv 0}  = - \frac{\theta}{2} e^{2r(T-t)} (x-y)^2 + e^{r(T-t) } x + \frac{\bmu^2}{2 \theta \sigma^2} (T-t).
\end{align}
The loss due to not taking insurance business is measured by
\begin{align}
\Vc(t,x,y) - \Vc(t,x,y)\Big|_{L \equiv 0} =  \frac{\left(\bp - \lambda \overline{\gamma}_1 - \frac{\rho \beta \bmu}{\sigma}  \right)^2}{\beta^2(1-\rho^2) + \lambda \overline{\gamma}_2 } \cdot \frac{T-t}{2 \theta}   > 0.
\end{align}
\end{corollary}

Several important remarks and explanations are due regarding Theorem \ref{thm:main} and Corollaries \ref{cor:main1} and \ref{cor:main2}.
First, the optimal strategy depends on both the financial market and the insurance market.
Namely, $\pi^*$ also depends on the risk model \eqref{eq:dR} and $\pi^* \neq 0$ in general, while $L^*$ depends on the price models \eqref{eq:dS0}-\eqref{eq:dS1} and $L^* \neq 0$ either. This observation testifies the importance of considering a combined financial and insurance market in the risk management study for an insurer.
Corollary \ref{cor:main1} further  shows that the loss in the value function is strictly positive for all $t \in [0,T)$;  not having the access to the financial market makes the insurer worse off.
On the other hand, when the insurer does not engage in the insurance business (i.e., $L = 0$), the insurer invests as if she were a utility maximizer equipped with an exponential utility  $U(x) = - e^{-\theta x}$.\footnote{
Please see Remark 1 in \cite{basak2010dynamic} for further discussions on the connection with exponential utility maximization. }
In this case, the loss in the value function is also strictly positive, as shown in Corollary \ref{cor:main2}, even when the insurance policy is ``cheap", i.e., when $\bp = \lambda \overline{\gamma}_1$ ($p \dd t = \Eb[\dd R(t)]$).
Along this discussion, given $\bp \le \lambda \overline{\gamma}_1$ (insurance policies are underpriced), the optimal risk control strategy $L^*$ may still be positive, if $\rho \beta  \bmu >0$ is big enough. Note that a necessary condition is $\rho > 0$, given $\bmu >0$.
This finding is certainly interesting, as the insurer takes positive shares in a ``losing" insurance business,
which seems an irrational decision at first thought.
But a second thought reveals that such a business provides a natural hedge to the risky asset, making it still desirable to hold positive shares in the portfolio.
Another plausible explanation is that the insurer sees underwriting policies as a financing tool to raise capital for investment purpose.
If the gain from investing premiums in the financial market outweighs the shortfall due to underpricing   policies, the insurer indeed has the incentive to sell underpriced insurance policies.

Next, we investigate the impact of the correlation between the two markets on the optimal strategy.
In the extreme case of $\rho = 0$, i.e., when the two markets are independent, we have
\begin{align}
\label{eq:op_zero}
\pi^*(s) \Big|_{\rho = 0} = \frac{\bmu  }{\theta \sigma^2 } \, e^{-r(T-s)} \qquad \text{ and } \qquad
& L^*(s) \Big|_{\rho = 0} = \frac{ \bp - \lambda \overline{\gamma}_1  }{\theta \left( \beta^2 + \lambda \overline{\gamma}_2 \right) } \, e^{-r(T-s)}, \qquad \forall \, s \in [t,T],
\end{align}
which implies the two markets are now disentangled, and investment and risk control decisions are independent.
Note that in this case the optimal investment strategy $\pi^* |_{\rho =0} = \pi^*|_{L = 0}$ is the same as the optimal strategy in the standard Merton's problem with an exponential utility $U(x) = - e^{-\theta x}$.
For the given risk process $R$ in \eqref{eq:dR}, if $p \le \alpha + \lambda \overline{\gamma}_1 $ (i.e., $\bp \le \lambda \overline{\gamma}_1$), ruin occurs for sure to the insurer. With that in mind, let us suppose the following conditions hold for the rest of the section:
\begin{align}
\label{eq:model_asu}
\bp - \lambda \overline{\gamma}_1 > 0 \qquad \text{ and } \qquad \bmu = \mu - r > 0,
\end{align}
where the second inequality means the risky asset has higher return than the risk-free asset.
Let us denote  $\pi^*|_{\rho > 0}$ the optimal investment in the risky asset under a positive correlation $\rho$.
We rewrite the optimal strategy in \eqref{eq:op} as follows
\begin{align}
\label{eq:op_decom}
\pi^*(s) = \frac{\beta^2 + \lambda \overline{\gamma}_2}{\beta^2 ( 1 - \rho^2) + \lambda \overline{\gamma}_2} \cdot \pi^*(s) \Big|_{\rho = 0} + \rho \, \frac{\beta(\bp - \lambda \overline{\gamma}_1)}{\left(\beta^2 ( 1 - \rho^2) + \lambda \overline{\gamma}_2 \right) \sigma} \, \frac{1}{\theta} \, e^{-r(T-s)},
\end{align}
from which we deduce $\pi^*(s)|_{\rho > 0}  > \pi^*(s)|_{\rho = 0}$.
The economic meaning is that an insurer, with risk positively correlated with the risky asset, invests more aggressively in the risky asset, as if she were less risk averse.
This makes perfect sense, since, with $\rho > 0$, a decrease in the price of the risky asset is accompanied by a decrease in the liabilities to be paid out, making the risky asset less risky to the insurer.
However, when these two markets are negatively correlated, less can be said, as although the factor in the first term in \eqref{eq:op_decom} is still greater than 1, the second term in \eqref{eq:op_decom} is negative.
In consequence, we expect different parameter values lead to  different monotonicity results.
One could carry out the same analysis on the optimal risk control strategy $L^*$, and the findings are the same.
We point out that the numerical analysis in \cite{zou2014optimal} shows $\pi^*$ is increasing with respect to $\rho \in (-1,0)$, but $L^*$ seems to exhibit a ``smile" shape (decreasing first and then increasing) as a function of negative $\rho$.

We continue to study the impact of other model parameters on the optimal strategy.
After careful computations and analysis, we summarize the impact of all the model parameters on the optimal strategy and the value function in Table \ref{tab:impact}. Note the results are obtained under the additional conditions in \eqref{eq:model_asu}.
As the excess return $\bmu$ (recall $\bmu = \mu -r$) increases, the optimal investment strategy $\pi^*$ increases, and the optimal liability strategy $L^*$ increases (resp. decreases) if and only if $\rho>0$ (resp. $\rho<0$).
The impact of the excess premium $\bp$ (recall $\bp = p - \alpha$) on the optimal strategy is exactly the opposite, comparing to that of $\bmu$.
However, how the rest of parameters affect the optimal strategy is less clear, with only partial results  as presented in Table \ref{tab:impact}. As already discussed in details in the proceeding paragraph, when $\rho > 0$, the financial and insurance markets provide a natural hedge to each other, which allows us to gain full insight on the optimal strategy.
When the two markets are negatively correlated, argued in \cite{stein2012stochastic} to be a main contributor to the failure of AIG during the financial crisis,  the change of a parameter (e.g., $\beta$ and  $\lambda$) results in opposite reactions from the two markets.
To better explain this result, let us consider the impact of the asset volatility $\sigma$ on the optimal investment $\pi^*$.
From \eqref{eq:op_decom}, it is clear that, as $\sigma$ increases, the myopic component $\bmu/ (\theta \sigma^2)$ decreases, but the second hedging component increases if $\rho < 0$.
In summary, the case of negative correlation $\rho < 0$ puts more uncertainty on the optimal decisions, and we will revisit this topic in numerical studies.

\begin{table}[h!]
\centering
\begin{tabular}{|c|c|c|c|}\hline
Parameter & Optimal Investment $\pi^*$ & Optimal Liability $L^*$ & Value Function $\Vc$ \\ \hline
correlation $\rho$ & $\partial \pi^*/ \partial \rho >0 $ if $\rho > 0$ & $\partial L^*/ \partial \rho >0 $  if $\rho > 0$
& $\partial \Vc / \partial \rho <0 $  if $\rho < 0$
\\ \hline
excess return $\bmu$ & $\partial \pi^*/ \partial \bmu >0 $ &  $\partial L^*/ \partial \bmu  = \text{ sign } (\rho)$
& $\partial \Vc / \partial \bmu >0 $  if $\rho < 0$
\\ \hline
asset volatility $\sigma$ & $\partial \pi^*/ \partial \sigma  <0 $ if $\rho > 0$ & $\partial L^*/ \partial \sigma  = \text{ sign } (\rho)$
& $\partial \Vc / \partial \sigma >0 $  if $\rho < 0$
\\ \hline
excess premium $\bp$ & $\partial \pi^*/ \partial \bp  = \text{ sign } (\rho)$   & $\partial L^*/ \partial \bp  >0$
& $\partial \Vc / \partial \bp <0 $  if $\rho < 0$
\\ \hline
liability volatility $\beta$ & $\partial \pi^*/ \partial \beta  >0 $ if $\rho > 0$  &  $\partial L^*/ \partial \beta >0 $  if $\rho = 0$
& $\partial \Vc / \partial \beta  <0 $  if $\rho < 0$
\\ \hline
jump intensity $\lambda$ & $\partial \pi^*/ \partial \lambda <0 $ if $\rho > 0$ & $\partial L^*/ \partial \lambda <0 $ if $\rho > 0$
& $\partial \Vc / \partial \lambda  <0 $  if $\rho > 0$
\\ \hline
jump size $\gamma$ & $\partial \pi^*/ \partial \gamma <0 $ if $\rho > 0$  & $\partial L^*/ \partial \gamma <0 $ if $\rho > 0$
& $\partial \Vc / \partial \gamma  <0 $  if $\rho > 0$
\\ \hline
\end{tabular}
\caption{Impact of Model Parameters on the Optimal  Strategy and the Value Function}
\label{tab:impact}
{\footnotesize Note. In the last row of ``jump size $\gamma$'', we derive the results assuming $\gamma(\cdot) \equiv \gamma$ is a positive constant.}
\end{table}

From the efficient frontier \eqref{eq:eff} of Problem \eqref{eq:prob2}, one immediate observation is the so-called security market  line (SML),
\begin{align}
	\Eb_{t,x} \left[X^*(s) \right]= x \, e^{r(s-t)}  + \sqrt{\kappa_3 (s-t) } \times \, \sqrt{\Vb_{t,x} \left[X^*(s)\right] }.
\end{align}
Since the dependence of the value function $\Vc(t,x,y)$ on the model parameters (except $r$) is only through the coefficient $\kappa_3$, the last column of Table \ref{tab:impact} offers some monotonic results regarding the slope of the SML.
For instance, when $\rho<0$, the slope of the SML is increasing with respect to (w.r.t.) the excess return $\bmu$ and asset volatility $\sigma$,  and decreasing w.r.t. the correlation coefficient $\rho$, the excess premium $\bp$, and the risk volatility $\beta$.

Finally, according to Theorem \ref{thm:main}, both the optimal investment strategy $\pi^*$ and the optimal liability strategy $L^*$ are \emph{independent} of the wealth, but are increasing at an exponential rate with respect to the time variable.
Setting $y=x$, we have $\Vc_x(t,x,x) >0$ and $\Vc_t(t,x,x)  < 0$.
That is, with higher initial wealth $x$ or a longer investment horizon $T-t$, the insurer is able to derive a higher expected utility, which fits our intuition perfectly.

\section{Comparison Analysis}
\label{sec:comp}

The goal of this section is to compare our approach and optimal strategy in Section \ref{sec:main} with those under the game-theoretic and precommitted framework.

\subsection{Comparison Analysis with Game-Theoretic Strategies}
\label{sub:comp_game}

\subsubsection{Comparison in Optimal Strategy}
If we consider a standard MV portfolio selection problem \emph{without} risk control strategies, the optimal investment $\pi^*|_{L \equiv 0}$ is obtained in \eqref{eq:pi_L=0}, which is the same as that in \cite{basak2010dynamic} (without stochastic factor), \cite{bjork2010general}, \cite{bjork2014mean} (under constant risk aversion), and \cite{kryger2019optimal}.
But as witnessed in Section \ref{sec:main}, we arrive at the same result via a different analysis.

Next, let us compare our optimal strategy in \eqref{eq:op} to those in the time-consistent MV investment-reinsurance literature.
If we set $\rho = 0$ and $\gamma \equiv 0$, i.e., the financial market is independent of the insurance market and the risk process $R$ is a diffusion approximation process, then we recover the same results as in \cite{zeng2011optimal}; see Theorem 2 therein. If we set $\rho = 0$, then our optimal strategy is consistent with that in \cite{zeng2013time} and \cite{zeng2016robust} without ambiguity aversion.\footnote{There is slight difference in the setup of risk management between ours and those in the works of Zeng and his collaborators'; see Remark \ref{rem:model}.
For instance, using our notation, $m $ and $\theta $ in \cite{zeng2011optimal} are 0 and $\bp - \lambda \gamma$.}

There is a key difference between our optimal strategy and those discussed above.
Under \eqref{eq:model_asu}, the optimal strategy in the above papers is always positive (excluding the wealth-dependent risk aversion case in \cite{bjork2014mean}) and does not involve short selling.
In our model, the optimal strategy is positive if $\rho \ge 0$.
However, if $\rho <0$,  it is possible that $\kappa_1$ and $\kappa_2$ are negative at the same time, or they  have different signs. In other words, our optimal strategy may involve short selling when $\rho < 0$; see Figure \ref{fig:rho_op}.

Our optimal strategy is independent of the wealth process $X$, since the risk aversion parameter $\theta$ is taken to be a constant in the analysis.
For the same problem but under a wealth-dependent risk aversion (e.g., $\theta(x) = \text{constant} / x $), the optimal strategy is likely to be proportional to the wealth level; see, \cite{bjork2014mean},  \cite{dai2020dynamic}, and \cite{kryger2019optimal}.

\subsubsection{Comparison in Equilibrium/Optimality Definition}
A standard approach to tackle the time-inconsistency issue of Problem \eqref{eq:prob} is to formulate the problem under a game-theoretic framework.
That is, one sets up a game between the current self of an MV-type agent and her future incarnations (selfs).
A strategy $\wh u$ is ``optimal", more often called equilibrium (used hereafter in this subsection), if all her future incarnations living in $[t+ \epsilon, T]$ will follow this strategy  and there is no gain for her to choose a different strategy $u$ at time $t$ (lasting from $t$ to $t + \epsilon$ for an infinitesimal period $\epsilon$).
The above ``informal" definition of an equilibrium strategy $\wh u$ is indeed rigorous in a discrete-time model.
To see this, suppose $\wh u = (\wh u(s))_{s \in [t,T]}$ is an equilibrium strategy and consider a ``perturbed" strategy $u=(u(s))_{s \in [t,T]}$, where $u(t) = u_0 \neq \wh u(t)$ and $u(s) = \wh u(s)$ for all $s=t+1, \cdots, T$.
By the definition of an equilibrium strategy, we have
$J(t,x; \wh u) \ge J(t,x; u)$
for any $\Fc_t$-measurable $u_0$ in the admissible domain.
However, extending the same idea to a continuous-time model becomes problematic, since a deviation from $\wh u$  at time $t$ is only effective for an infinitesimal period of time (a set with Lebesgue measure zero) and hence its impact on the objective functional is negligible in most important problems (including the MV problems).
To overcome this issue, a modified (and weakened) equilibrium condition is introduced.
In the continuous-time framework, a strategy $\wh u$ is called an equilibrium strategy if
\begin{align}
\label{eq:equi_cond}
\liminf_{\epsilon \downarrow 0} \, \frac{J(t,x; \wh u) - J(t,x; u)}{\epsilon} \ge 0,
\end{align}
holds for all perturbed strategies $u$ within the admissible set,
where we assume the goal is to maximize the objective $J$ in the problem.
The definition \eqref{eq:equi_cond} is proposed by \cite{ekeland2006being} to study an optimization problem with a non-exponential discounting function.
The equilibrium definition \eqref{eq:equi_cond} is also fundamental and used in almost all the subsequent works on time-consistent MV and MV-reinsurance problems; see, e.g., \cite{bjork2010general}, \cite{kryger2010some}, \cite{zeng2011optimal}, \cite{zeng2013time}, \cite{bjork2014mean}, and many others.

However, the equilibrium condition \eqref{eq:equi_cond} is only a \emph{necessary} condition, \emph{not} a sufficient condition.
That brings an immediate problem: a strategy $\wh u$ satisfying \eqref{eq:equi_cond} with an equality may fail to dominate another admissible strategy (See Remark 7.1 in \cite{bjork2010general}).
\cite{huang2020strong} construct a counterexample (see Example 4.3 therein) in which an equilibrium strategy $\wh u$ satisfying \eqref{eq:equi_cond} is strictly worse off than a different admissible strategy $u'$, i.e.,  $J(t,x; u') > J(t,x; \wh u)$.
Namely, there exist cases where an agent has the incentive  to deviate from an equilibrium strategy $\wh u$, as characterized by \eqref{eq:equi_cond}, which contradicts the very definition of equilibrium.

In comparison, the problem considered in this paper, Problem \eqref{eq:prob2}, is a standard time-consistent control problem. Namely, once a solution $u^*$ is obtained to Problem \eqref{eq:prob2} at time $t$, the insurer will follow the strategy $u^*$ over $[t,T]$ and the optimality of $u^*$ holds trivially by definition, i.e.,
\begin{align}
\label{eq:op_cond}
\Jc(t,x,y ; u^*) \ge \Jc(t,x,y; u) \qquad \forall \, u \in \Ac.
\end{align}
In fact, our analysis in Section \ref{sec:main} does find an optimal strategy $u^*$ such that $\Jc(t,x,y;u^*)  = \sup_{u \in \Ac} \, \Jc(t,x,y)$; see Theorem \ref{thm:main} and \eqref{eq:veri}.
The key differences between the equilibrium definition of \cite{bjork2010general} in \eqref{eq:equi_cond} and the optimality definition in \eqref{eq:op_cond} are as follows: (1) we introduce an auxiliary process $Y$ (with initial value $y$) and consider a modified objective $\Jc$, in which $Y$ replaces the conditional expectation in the original objective $J$; and (2) we seek an optimal strategy $u^*$ that maximizes $\Jc$ over all admissible strategies.
Because of these differences, our alternative formulation in \eqref{eq:prob2} always leads to a well-defined optimal  strategy (for a modified objective $\Jc$), while the same cannot be said in general under the definition \eqref{eq:equi_cond}.

\subsubsection{Comparison in Approach}

We discuss two differences between our approach and those in \cite{bjork2010general} and its following works, such as \cite{zeng2011optimal} and \cite{bjork2014mean}.
The first one comes from the auxiliary process, while the second one comes from the HJB equation.

To facilitate the presentation, let us restate the MV problem considered in \cite{bjork2010general} and \cite{bjork2014mean} as follows:
\begin{align}
\wh V(t,x) = \sup_{u \in \Ac} \; \wh J(t,x; u) = \sup_{u \in \Ac} \left\{ \Eb_{t,x} \left[X^u(T)\right] - \frac{\theta}{2} \Vb_{t,x} \left[X^u(T)\right] \right\}, \qquad \theta > 0.
\end{align}
In  \cite{bjork2010general},
an essential technique to handle the ``troubling" square term is to introduce an auxiliary process $g^u$ defined by
\begin{align}
\label{eq:g}
g^u(t, x) = \Eb_{t, x} \left[ X^u(T) \right], \quad t < T \qquad \text{ and } \qquad g^u(T, x) = x.
\end{align}
Notice that $g^u$ in \eqref{eq:g} is defined in a \emph{backward} way and its dynamics $g^u(t, X^u(t))$ are \emph{stochastic}, while our auxiliary process $Y^u$ is defined in a \emph{forward} way and its dynamics $Y^u(t)$ are \emph{deterministic}; see \eqref{eq:dY}.
Due to the deterministic nature of process $Y^u$, we do not have  $\Vc_{yy}$ and $\Vc_{xy}$ terms in our HJB equation \eqref{eq:hjb}.

For any admissible control $u \in \Ac$, define the associated Dynkin operator of process $X^u$ in \eqref{eq:dX} by $\Lc^u$.
As $u \in \Ac$, $X^u$ is the unique strong solution to the SDE \eqref{eq:dX} and $X(T)$ is square integrable.
It then follows from  \eqref{eq:g} that $g^u(t, X^u(t))$ is a martingale and we have $\Lc^u \, g^u(t, x) = 0$ for all $u \in \Ac$ by Dynkin's formula.
Assume an  equilibrium strategy $\hat u$ as defined in \eqref{eq:equi_cond} exists, \cite{bjork2010general} derive an extended system of HJB equations satisfied by the pair $(\wh V(t,x), \, g^{\hat u}(t,x))$ and solve the system to obtain $\hat u$; see Theorems 2.1 and 7.4 therein.
In our formulation, Problem \eqref{eq:prob2} is a standard control problem, with two state variables, and the HJB equation \eqref{eq:hjb} is about the value function $\Vc(t,x,y)$ only.
To summarize, the approach of \cite{bjork2010general} ends up with solving a system of two one-dimensional partial differential equations (PDEs), while our approach leads to a two-dimensional PDE.

\subsection{Comparison Analysis with Precommitment Strategies}
\label{sub:pre}

In Section \ref{sec:main}, we obtain the precommitment strategy $u^\pre$ to Problem \eqref{eq:prob} in \eqref{eq:pre_op2} and the optimal strategy $u^*$ to Problem \eqref{eq:prob2} in \eqref{eq:op}.
Since $u^\pre$ and $u^*$ are solutions to two different stochastic control problems, a direct side-by-side comparison has little mathematical meaning.
However, Problem \eqref{eq:prob2} is an alternative time-consistent formulation to Problem \eqref{eq:prob}, and both $u^\pre$ and $u^*$ are available investment and risk control strategies to the insurer, comparing the end results of these strategies makes economic sense.

First, we fix the same risk aversion parameter $\theta$ for both Problems \eqref{eq:prob} and \eqref{eq:prob2}, and investigate three important results: the mean $\Eb_{t,x}[X^u(T)]$, the variance $\Vb_{t,x}[X^u(T)]$, and the objective $J$ defined in \eqref{eq:obj}.
Let $X^\pre$ and $X^*$ denote the insurer's wealth process \eqref{eq:dX} under the precommitment strategy $u^\pre$ and the optimal strategy $u^*$, respectively.	
By comparing \eqref{eq:pre2_EV}-\eqref{eq:mv-pre} with \eqref{eq:op_EW}-\eqref{eq:op_VW} and \eqref{eq:mv-dyma}, we obtain:
\begin{align}
	\label{eq:comp1.1}
\Eb_{t,x}[X^\pre(T)]> \Eb_{t,x}\left[X^*(T)\right], \quad  \Vb_{t,x} [X^\pre(T)] > \Vb_{t,x} \left[X^*(T)\right], \quad J(t,x; u^\pre) > J(t,x; u^*) .
\end{align}
From \eqref{eq:comp1.1}, we conclude that the optimal strategy $u^*$ is more conservative than the precommitment strategy $u^\pre$, leading to a smaller risk under the compromise of performance (mean).
Note that to derive \eqref{eq:comp1.1}, we take the parameter $\theta$ to be the same for Problems \eqref{eq:prob} and \eqref{eq:prob2}. However, the results in \eqref{eq:comp1.1} clearly reveal different risk attitudes in terms of both mean and variance.
As such, in the next step, we fix the same target $\Eb_{t,x}[X^u(T)]$ for the insurer and study how the two different formulations achieve the same target.

Second, let us fix $\mb:= \Eb_{t,x} [X^u(T)]> x e^{r(T-t)}$ for the insurer, and re-consider Problems \eqref{eq:prob} and \eqref{eq:prob2} under a constrained admissible set $\Ac_{\mathbf{m}}$, defined by
\begin{align}
	\label{eq:Ac_m}
	\Ac_{\mathbf{m}} := \{ u \in \Ac : \Eb_{t,x} [X^u(T)] = \mathbf{m}\}.
\end{align}
Denote the corresponding solutions by $u^\pre_{\mb}$ and $u^*_{\mb}$.	We apply Theorems \ref{thm:pre} and \ref{thm:main} to obtain $u^\pre_{\mb}$ and $u^*_{\mb}$ in the next two corollaries and then compare them.

\begin{corollary}
	\label{cor:pre2}
	Let a target level $\mb := \Eb_{t,x} [X^u(T)]> x e^{r(T-t)}$ be given.
	A precommitment strategy, denoted by $u^\pre_\mb = (\pi^\pre_\mb(s), L^\pre_\mb(s))_{s \in [t,T)}$, to Problem \eqref{eq:prob} over the admissible set $\Ac_\mb$ is given by
	\begin{align}
		\label{eq:pre_m}
		\pi^\pre_\mb(s) &= - \kappa_1 \left(X^\pre_{\mb}(s) -  \frac{\mb - x \, e^{(r - \kappa_3) (T-t)}}{1 - e^{-\kappa_3 (T-t)}}\, e^{-r(T-s)}\right) \qquad \text{ and } \qquad L^\pre_\mb(s) = - \frac{\kappa_2}{\kappa_1} \, \pi^\pre_\mb (s) ,
	\end{align}
	where $X^\pre_{\mb}$ denotes the wealth process under the precommitment strategy $u^\pre_\mb$.
\end{corollary}

\begin{proof}
Recall $\Eb_{t,x}[X^\pre(T)]$ is obtained in \eqref{eq:pre2_EV}. By equating $\Eb_{t,x}[X^\pre(T)] = \mb$, we get
	\begin{align}
	\theta^\pre(\mb) = \frac{e^{\kappa_3(T-t) - 1} } {\mb -  x e^{r(T-t)}}  > 0.
	\end{align}	
		 Substituting the free parameter $\theta$ by the above $\theta^\pre(\mb)$  in \eqref{eq:pre_op2} leads to the desired results in \eqref{eq:pre_m}.
\end{proof}

\begin{corollary}
	\label{cor:pre}
	Let a target level $\mb := \Eb_{t,x} [X^u(T)]> x e^{r(T-t)}$ be given.
	A precommitment strategy, $u^*_\mb = (\pi^*_\mb(s), L^*_\mb(s))_{s \in [t,T)}$, to Problem \eqref{eq:prob2} over the admissible set $\Ac_\mb$ is given by 
	\begin{align}
		\label{eq:pre_op}
		\pi^*_\mb(s) = \frac{\kappa_1 \left(\mb - x \, e^{r(T-t)}\right)}{\kappa_3 (T-t)} \, e^{-r(T-s)} \qquad \text{ and } \qquad
		L^*_\mb(s) = \frac{\kappa_2 \left(\mb - x \, e^{r(T-t)}\right)}{\kappa_3 (T-t)} \, e^{-r(T-s)} .
	\end{align}
\end{corollary}

\begin{proof}
Recall $\Eb_{t,x} \left[X^*(s)\right]$ is obtained in \eqref{eq:op_EW}. By equating $\Eb_{t,x} [X^*(T)] = \mathbf{m}$ , we get
	\begin{align}
	\theta^*(\mb) = \frac{\kappa_3(T-t) } {\mb -  x e^{r(T-t)}}  > 0.
\end{align}	
 Substituting the free parameter $\theta$ by the above $\theta^*(\mb)$ in \eqref{eq:op} leads to the desired results in \eqref{eq:pre_op}.
\end{proof}

Despite successfully obtaining $\pi^\pre_{\mb}$ and $\pi^*_{\mb}$ in close-form, we cannot directly compare them at an arbitrary time $s$ ($t \le s <T$). However, when $s = t$, by using \eqref{eq:pre_m} and \eqref{eq:pre_op}, and $e^{\kappa_3(T-t)} - 1 > \kappa_3(T-t)$, we obtain
	\begin{align}
	\label{eq:comp2}
	\pi^\pre_{\mb} (t) > \pi^*_{\mb} (t) \qquad \text{and} \qquad L^\pre_{\mb}(t) > L^*_{\mb}(t).
	\end{align}
\eqref{eq:comp2} implies that, to achieve the same target $\mb$, the precommitment framework yields a more risky strategy in both investment and liability at the initial time $t$ than the alternative time-consistent formulation.

Note that we call $u^*_{\mb}$ in Corollary \ref{cor:pre2} a \emph{precommitment} strategy, since both $\pi^*_{\mb}$ and $L^*_{\mb}$ depend on the initial state $(t,x)$ as seen in \eqref{eq:pre_op}.
That means, by restricting the admissible set from $\Ac$ to $\Ac_\mb$, the alternative time-consistent formulation in Problem \eqref{eq:prob2} becomes time-inconsistent.
In fact, $u^*_\mb$ being a precommitment strategy is not surprising at all, because  the restriction of $\Ac_\mb$ leads to a state-dependent risk aversion $\theta^*(\mb)$, which is a well-known contributor to time-inconsistent problems  (see \cite{bjork2014mean}).

\section{Numerical Studies}
\label{sec:nume}

Given the explicit results on both the optimal strategy and the value function in Theorem \ref{thm:main}, we discuss their analytical properties with respect to the model parameters, and summarize the main findings in Table \ref{tab:impact}.
However, most findings there are partial and limited to the assumption of $\rho > 0$.
That motivates us to conduct numerical studies in this section to further investigate how the model parameters and risk aversion affect the insurer's decisions.

We follow \cite{zou2014optimal} to  initiate the default model parameters in \eqref{eq:dS0}-\eqref{eq:dR}, and present them in Table \ref{tab:model}, where we assume $\gamma(t,z) \equiv \gamma$ for all $t \in [0,T]$ and $z \in \Rb$.  Note that we leave both $\rho$ (correlation coefficient) and $\theta$ (risk aversion) unspecified in the default setting, as they are the most significant factors of the optimal strategy.
\begin{table}[htb]
\centering
\begin{tabular}{cccccccc} \hline
$r$ & $\mu$ &  $\sigma$ & $\alpha$ & $\beta$ & $\lambda$ & $\gamma$ & $p$ \\ \hline
0.01 & 0.05 & 0.25 & 0.08 & 0.1 & 0.1 & 0.3 & 0.15 \\ \hline
\end{tabular}
\caption{Default Model Parameters in \eqref{eq:dS0}-\eqref{eq:dR}}
\label{tab:model}
\end{table}

\begin{figure}[htb]
	\begin{center}
		\includegraphics[width = \textwidth]{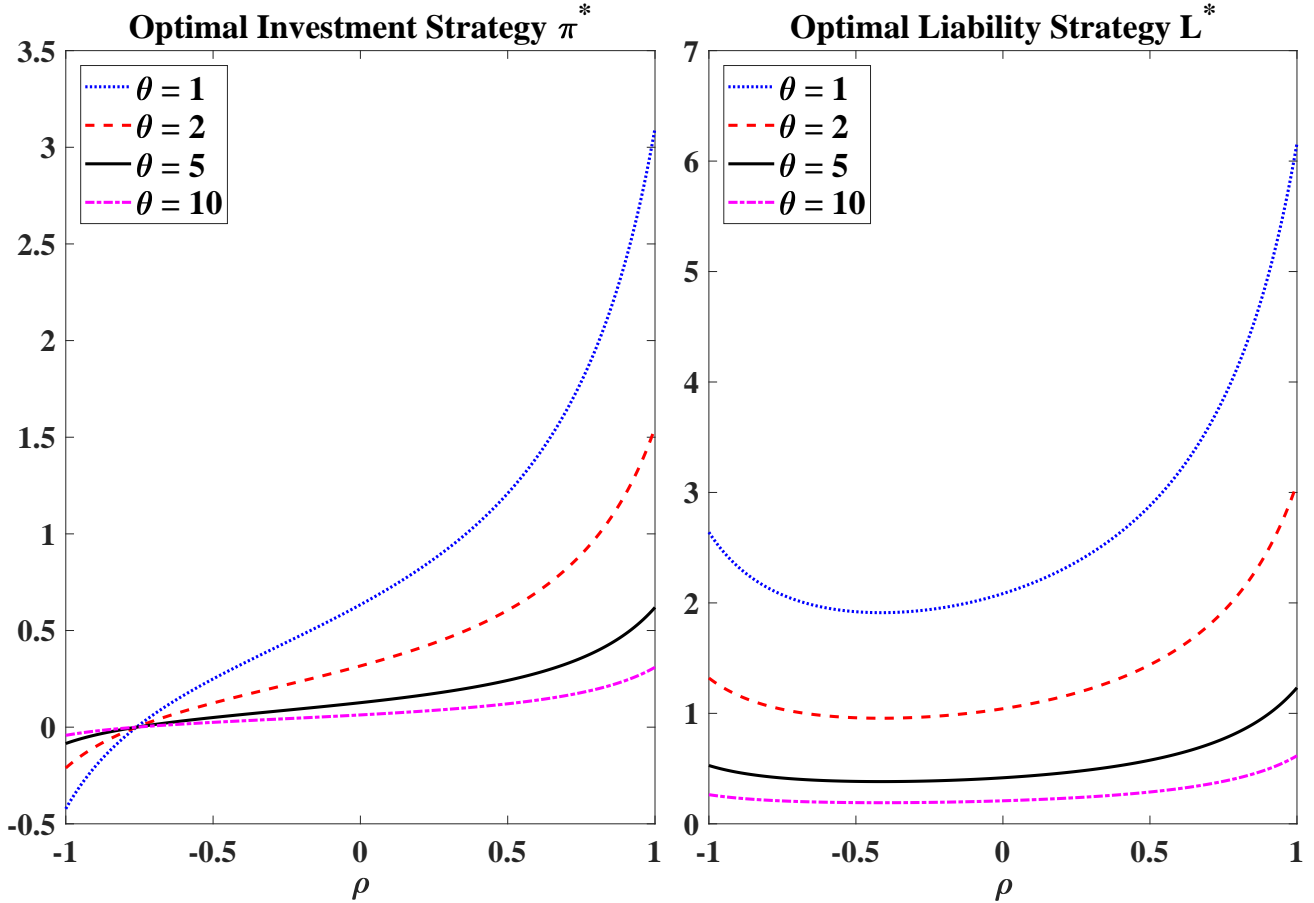}\\[-4ex]
		\caption{Impact of $\rho$ and $\theta$ on the Optimal Strategies $\pi^*$ (left) and $L^*$ (right)}
		\label{fig:rho_op}
	\end{center}
\centering
	\vspace{-2ex}
	{\small Notes. We plot $\pi^*(t)$ and $L^*(t)$  under the default model parameters in Table \ref{tab:model}, $x=1$, and  $T-t =1$.}
\end{figure}

First, we examine the impact of $\rho$ (correlation coefficient) and $\theta$ (risk aversion) on the optimal strategy. We plot the optimal investment strategy $\pi^*(t)$ and the optimal liability strategy $L^*(t)$ against all $\rho \in [-1,1]$ under four different risk aversion levels $\theta = 1, 2, 5, 10$ in Figure \ref{fig:rho_op}.
Several important observations and explanations are due as follows.
\begin{itemize}
	\item The optimal investment in the risky asset may be negative (i.e., shorting selling may be optimal), but only when $\rho$ is close to -1. We further observe that a less risk averse insurer  short sells more in a combined market with extreme negative correlation. To understand these results, we look at the  extreme case of $\rho = -1$, in which the movements from the Brownian motion $W_1$ cause the risky asset $S_1$ and the risk process $R$ to affect the insurer in the exactly \emph{same} direction, if she holds positive positions.
	For instance, a decrease of $W_1$ leads to a lower price of $S_1$ and more liabilities  from underwriting policies.
	 In turn, the volatility due to $W_1$ is amplified and results in a more volatile market in the insurer's view.
	A strategy to counter such an amplification effect is to short sell the risky asset.
	A less risk averse insurer is less prone to the amplification effect, and then short sells more risky asset when $\rho$ is close to -1.
	
	\item The optimal investment $\pi^*$ is an increasing function of $\rho$, and shares the same intersection point at $\pi^*=0$ (recall $\pi^*=0$ iff $\kappa_1=0$ by \eqref{eq:op} and $\kappa_1$ is independent of $\theta$ by \eqref{eq:kappa12}).
	The optimal liability $L^*$ first decreases and then increases with respect to (w.r.t.) $\rho$, revealing a convex relation.
	However, if $\rho > 0$, $L^*$ is an increasing function of $\rho$, as already shown in Table \ref{tab:impact}.
	Under the given parameters in Table \ref{tab:model}, we calculate that $L^*$ takes minimum values at $\rho = - 0.4143$.
	These findings are consistent with those in \cite{zou2014optimal}.
	As already explained in Section \ref{sec:main}, when $\rho > 0$, the financial market and the insurance market provide a natural hedge to each other, making the combined market less volatile to the insurer, and hence, both $\pi^*$ and $L^*$ increase w.r.t. $\rho$ when $\rho > 0$.
	
	\item As risk aversion $\theta$ increases, we observe that the optimal liability strategy $L^*$ taken by the insurer decreases, and is almost insensitive to $\rho$ when $\theta$ is large enough.
	Similar statement holds true for the optimal investment $\pi^*$ (in the absolute value).

\end{itemize}

\begin{figure}[htb]
	\begin{center}
		\includegraphics[width = \textwidth]{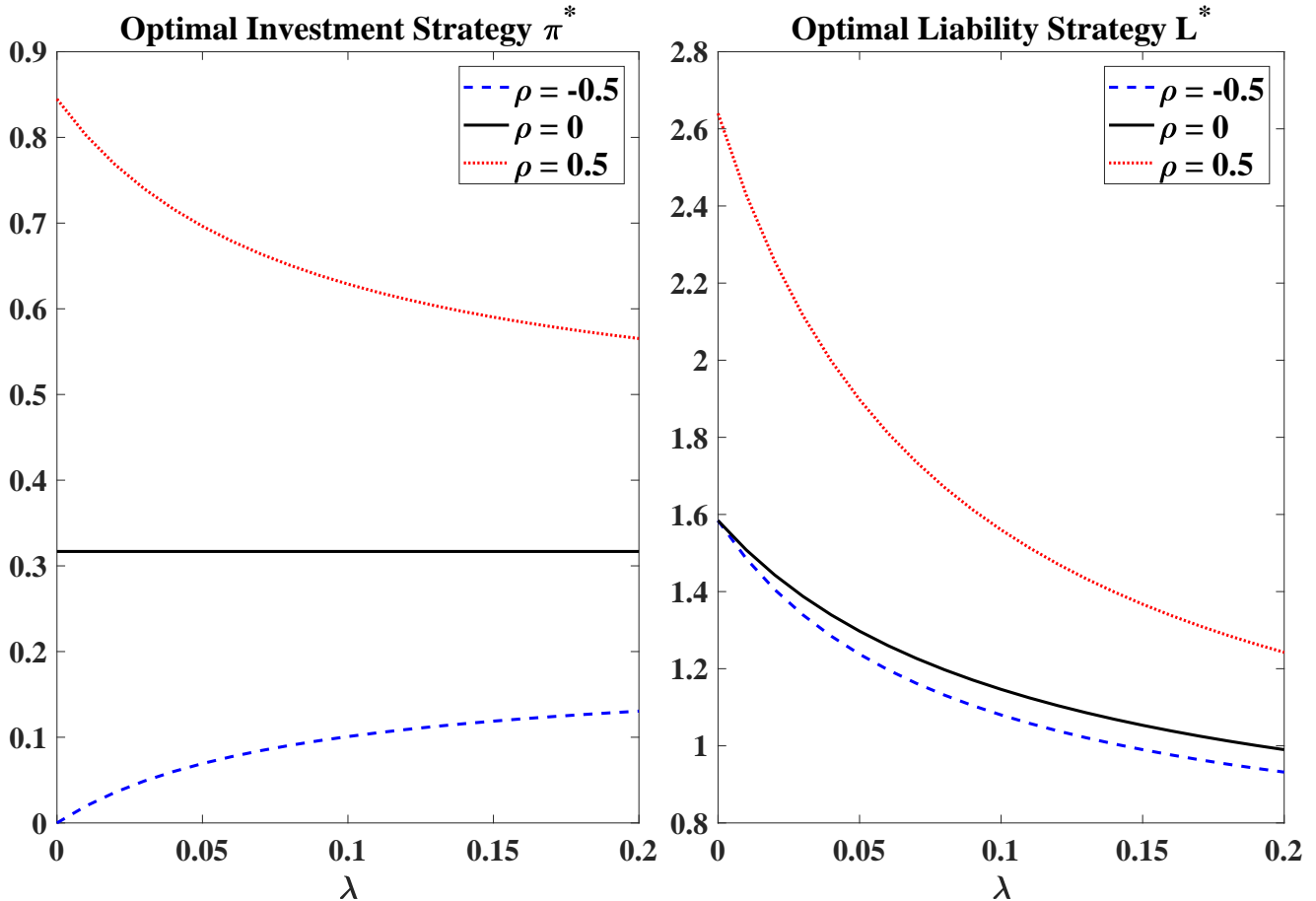}\\[-4ex]
		\caption{Impact of $\lambda$ on the Optimal Strategies $\pi^*$ (left) and $L^*$ (right)}
		\label{fig:lam_op}
	\end{center}
	\vspace{-2ex}
	{\small Notes. We calculate the premium $p$ by the expected value principle with loading $\eta = 40\%$. We set $\theta = 2$, $x=1$, $T-t =1$, 	and other parameters as in Table \ref{tab:model}.}
\end{figure}

We next study how the jump intensity $\lambda$ of the risk process $R$ affects the insurer's optimal strategy.
To this end, we allow $\lambda$ to vary in $[0, 0.2]$. We no longer fix $p=0.15$ as now $\lambda$ is changing, but instead apply the expected value principle with loading $\eta = 40\%$ to calculate the premium $p$ by $p = (1 + \eta) \times (\alpha + \lambda \gamma)$. We set $\theta = 2$ and consider three correlation levels $\rho = -0.5, 0, 0.5$. The remaining model parameters are the same as in Table \ref{tab:model}.
We then plot $\pi^*(t)$ and $L^*(t)$, with $x=1$ and $T-t = 1$, against $\lambda \in [0, 0.2]$ in Figure \ref{fig:lam_op}.
We comment that the plots are consistent with the partial findings on $\lambda$ in Table \ref{tab:impact}.
When $\rho = 0$, the flat solid line (in black) in the left panel of Figure \ref{fig:lam_op} represents the Merton ratio.
The impact of $\lambda$ on $L^*$ is more direct and easy to understand. As $\lambda$ increases, the liabilities per unit increase and the insurer responds by taking less units (policies) in the insurance business.
However, the impact of $\lambda$ on $\pi^*$ is indirect and more complex.
We notice a decreasing  relation  when $\rho = 0.5$ and an increasing relation   when $\rho = -0.5$.
In a positively correlated combined market, the risk process $R$ acts as a hedge to the risky asset.
But, as $\lambda$ increases, this hedging effect weakens; or putting it differently, the hedging tool itself becomes too volatile to serve its hedging purpose.
Upon understanding that, the insurer reacts by reducing risky investment.
Less can be said in general regarding the relation between $\pi^*$ and $\lambda$ when $\rho$ is negative.
We know that, with $\rho < 0$,  the increase of $\lambda$ leads to a shrinking effect on the risky asset's volatility $\sigma$, making the insurer willing to hold more risky asset, but at the same time a bigger $\lambda$ makes the insurance business more volatile, causing the insurer to be cautious of making risky investment.
The former effect outweighs the latter one, leading to a positive relation found in Figure \ref{fig:lam_op}.
We remark that the opposite may happen if a different set of model parameters is taken.

\section{Conclusions}
\label{sec:conl}

We study an optimal investment and risk control optimization problem for an insurer with mean-variance (MV) preference.
The insurer has access to a combined market with financial assets and insurance business opportunities.
The financial market is given by a standard Black-Scholes model, and the risk process (liabilities) per unit is given by a jump-diffusion process.
The insurer seeks an optimal investment and risk control strategy under the MV preference, which is a well-known time-inconsistent problem.
We introduce a deterministic auxiliary process to replicate the conditional expectation of the insurer's wealth and extend it as a second state process.
We then formulate an alternative time-consistent problem which is not only intimately related to the original problem but also can be solved by the standard dynamic programming method.
We obtain the optimal strategy, the efficient frontier, and the value function, all in closed-form, to the alternative problem.
We conduct  analytical studies to compare our formulation and optimal strategy with those under the  game-theoretic and precommitted framework.
Among many findings, we emphasize that the correlation between the financial market and the risk process plays a key role in the optimal strategy.
When the correlation coefficient $\rho$ is positive, both the optimal investment strategy $\pi^*$ and the optimal liability strategy $L^*$ are increasing functions of $\rho$.
However, when $\rho$ is negative, $L^*$ first decreases and then increases with respect to $\rho$, showing a convex U-shaped relation.
As $\rho$ moves towards -1, $\pi^*$ becomes negative, suggesting the optimal investment decision is to short sell the risky asset.
Lastly, we report that the jump intensity of the risk process also has a major impact on the optimal strategy.

\section*{Acknowledgments}

We would like to thank anonymous referees and editor for their careful reading and many insightful comments that help us improve the quality of an early version of this paper.
Bin Zou is partially supported by a start-up grant from the University of Connecticut.
Yang Shen is partially supported by the Discovery Early Career Researcher
Award (Grant No. DE200101266) from the Australian Research
Council.

\vspace{2ex}
\noindent
Declarations of interest: none.

\appendix

\section{Proofs}
\label{app:pre}

Problem \eqref{eq:prob} is a standard MV type control problem that is well studied in the literature; see, e.g., \cite{markowitz1952portfolio} and \cite{li2000optimal} in discrete time,  and \cite{zhou2000continuous} and the book \cite{yong1999stochastic} in continuous time.
There are various approaches to finding the precommitment strategy, and most of them rely on the so-called embedding technique, first used in \cite{li2000optimal} and \cite{zhou2000continuous}.
We denote a constrained admissible set $\Ac_{\mathbf{m}}$ by
\begin{align}
	\label{eq:Ac_m1}
	\Ac_{\mathbf{m}} := \{ u \in \Ac : \Eb_{t,x} [X^u(T)] = \mathbf{m}\}, \quad \text{where } \mb > x \, e^{r(T-t)},
\end{align}
and explain such a technique using the (equivalence) relations as follows:
\begin{align}
\sup_{u \in \Ac} \, J(t,x;u)\quad &\Leftrightarrow  \quad \sup_{\mb \in \Rb} \, \sup_{u \in \Ac_\mb} \, J(t,x;u) = \mb - \frac{\theta}{2} \Eb_{t,x} \left[ \left(X^u(T) - \mb \right)^2 \right]
\quad  \Leftrightarrow \quad \inf_{u \in \Ac_\mb} \, \Vb_{t,x} \left[ X^u(T) \right] \\
& \Leftrightarrow \quad \inf_{u \in \Ac} \, \Vb_{t,x} \left[ X^u(T) \right] - 2 \mathbf{w} \left( \Eb_{t,x} \left[ X^u(T) \right] - \mb \right) \\
& \Leftrightarrow \quad  \inf_{u \in \Ac} \, \Eb_{t,x} \left[ \left( X^u(T)  - \xi \right)^2\right] - (\xi - \mb)^2, \qquad \xi = \mb + \mathbf{w} \text{ (Lagrange multiplier)}.
\label{eq:road}
\end{align}
The above relation shows that the key to solving Problem \eqref{eq:prob} is finding an optimal solution to the following problem:
\begin{align}
\label{eq:aux_prob}
\Wc(t,x; \xi) := \inf_{u \in \Ac} \, \Eb_{t,x} \left[ \left( X^u(T)  - \xi \right)^2\right], \qquad \text{ where } \xi \in \Rb.
\end{align}

\begin{theorem}
	\label{thm:aux}
Suppose $\beta^2(1-\rho^2) + \lambda \overline{\gamma}_2  \neq 0$. An optimal solution to Problem \eqref{eq:aux_prob}  is given by
\begin{align}
\label{eq:aux_op} \quad
\pi^\pre(s; \xi) = - \kappa_1 \left(X^\pre(s;\xi) - \xi \, e^{-r(T-s)}\right) \quad \text{ and } \quad
L^\pre(s; \xi) = - \kappa_2 \left(X^\pre(s; \xi) - \xi \, e^{-r(T-s)}\right),
\end{align}
where $\kappa_1$ and $\kappa_2$ are defined in \eqref{eq:kappa12} and $X^\pre(\cdot;\xi)$ is the corresponding wealth process under the initial state $(t,x)$.
The value function to Problem \eqref{eq:aux_prob} is given by
\begin{align}
\label{eq:aux_value}
\Wc(t,x; \xi) = \left(x \, e^{r(T-t)} - \xi \right)^2 \, e^{-\kappa_3 (T-t)},
\end{align}
where $\kappa_3$ is defined in \eqref{eq:kappa3}.
\end{theorem}

\begin{proof}
Since Problem \eqref{eq:aux_prob} is a standard control problem, by using the dynamic programming principle, we obtain that $\Wc$ solves the following HJB equation (assuming $\Wc$ satisfies the needed regularity conditions)
\begin{align}
\Wc_t(t,x; \xi) + \sup_{(\pi, L) \in \Rb^2} \, \bigg\{ (rx + \bmu \pi + \bp L) \, \Wc_x(t,x; \xi)  + \frac{1}{2} \left(\sigma^2 \pi^2 - 2 \rho \beta \sigma \pi L + \beta^2 L^2 \right) \Wc_{xx}(t,x; \xi) \\
+ \lambda \, \int_{\Rb} \, \big( \Wc(t, x - L \gamma(z); \xi) - \Wc(t,x; \xi) \big) \, \dd F_Z(z) \bigg\} = 0
\end{align}
along with the terminal condition $\Wc(T,x; \xi) = (x - \xi)^2$.
We then make an educated guess for the value function in the form of
\begin{align}
\Wc(t,x ; \xi) = \left(x \, e^{r(T-t)} - \xi \right)^2 \, f(t), \qquad f(T) = 1.
\end{align}
Straightforward computations and verification then yield the desired results in \eqref{eq:aux_op} and \eqref{eq:aux_value}.
\end{proof}

We next apply the above theorem and the road map in \eqref{eq:road} to solve Problem \eqref{eq:prob}.

\begin{proof}[Proof of Theorem \ref{thm:pre}]
Using \eqref{eq:aux_value}, we consider the problem below
\begin{align}
\wh \Wc(t,x; \mb) =  \sup_{\xi \in \Rb} \, \Wc(t,x; \xi) - (\xi - \mb)^2
\end{align}
and easily obtain the optimal solution and the value function, under the given expectation level $\mb$, by
\begin{align}
\label{eq:xi_op}
\xi^*(\mb) = \frac{\mb - x \, e^{(r - \kappa_3) (T-t)}}{1 - e^{-\kappa_3 (T-t)}} \qquad \text{ and } \qquad
\wh \Wc(t,x; \mb) = \frac{\left(\mb - x \, e^{r (T-t)} \right)^2}{e^{\kappa_3 (T-t)} - 1}.
\end{align}
Next, we solve
\begin{align}
\sup_{\mb \in \Rb} \; \mb - \frac{\theta}{2}\wh \Wc(t,x; \mb)
\end{align}
and obtain the optimal target level $\mb^*$ by
\begin{align}
\label{eq:mb_op}
\mb^* = x \, e^{r(T-t)} + \frac{e^{\kappa_3 (T-t)} - 1}{\theta}.
\end{align}
Finally, we substitute $\xi$ in \eqref{eq:aux_op} by $\xi^*(\mb^*)$, where $\xi^*(\cdot)$ and $\mb^*$ are given in \eqref{eq:xi_op} and \eqref{eq:mb_op}, and obtain the desired results after tedious computations.
\end{proof}

\begin{proof}[Proof to Theorem \ref{thm:main}]
	From the terminal condition \eqref{eq:bound} and the nature of stochastic linear-quadratic problems, we guess an ansatz to the value function in the form of
	\begin{align}
		\Vc(t,x,y) = A(t) \, (x-y)^2 + B(t) \, x + C(t),
	\end{align}
	where $A$, $B$ and $C$ are yet to be determined and $A(t) < 0$ for all $t$.
	Note that with $A(t) < 0$, the above $\Vc$ is concave in both $x$ and $y$ arguments.
	It is clear from \eqref{eq:bound} that
	\begin{align}
		A(T) = - \frac{\theta}{2}, \qquad B(T) = 1, \qquad C(T) = 0.
	\end{align}
	
	Using the above ansatz and the HJB equation \eqref{eq:hjb}, we derive that an (candidate of) optimal strategy $u^* = (\pi^*, L^*)$  should solve the following system of equations
	\begin{align}
		\begin{cases}
			\sigma^2 \cdot \pi^* - \rho \beta \sigma \cdot  L^* = - \dfrac{\bmu \, B(t) }{2 A(t) }\\
			\rho \beta \sigma \cdot \pi^* - \left(\beta^2 + \lambda \overline{\gamma}_2 \right) \cdot L^* = \dfrac{(\bp - \lambda \overline{\gamma}_1 ) B(t)}{2 A(t)}
		\end{cases}
	\end{align}
	which, under the assumption $\beta^2(1-\rho^2) + \lambda \overline{\gamma}_2  \neq 0$ in \eqref{eq:assu}, admits a unique solution
	\begin{align}
		\pi^* = - \kappa_1 \, \frac{B(t)}{2 A(t) } \qquad \text{ and } \qquad L^* = - \kappa_2 \frac{B(t)}{2 A(t)},
	\end{align}
	where $\kappa_1$ and $\kappa_2$ are defined in \eqref{eq:kappa12}.

	We plug the above $(\pi^*, L^*)$ into the HJB equation \eqref{eq:hjb} and simplify to get
	\begin{align}
		(x-y)^2 A'(t) + x B'(t) + C'(t) + 2r(x-y)^2 A(t) + r  x B(t) - \kappa_3 \frac{B^2(t)}{4 A(t)} = 0,
	\end{align}
	where $\kappa_3$ is defined by \eqref{eq:kappa3}.
	Since the above equation holds for all $x, y \in \Rb$, we obtain
	\begin{align}
		\begin{cases}
			A'(t) + 2r \, A(t) =0,  & A(T) = - \frac{\theta}{2}, \\
			B'(t) + r \, B(t) = 0, & B(T) = 1, \\
			C'(t) - \kappa_3 \, \frac{B^2(t)}{4 A(t)} = 0,  & C(T) = 0,
		\end{cases}
	\end{align}
	which leads to the solutions
	\begin{align}
		A(t) = - \frac{\theta}{2} e^{2r(T-t)}<0, \qquad B(t) = e^{r(T-t)}>0, \qquad  C(t) = \frac{\kappa_3}{2 \theta} (T-t)>0.
	\end{align}
	It is straightforward to verify that $\Vc$ given by \eqref{eq:value} is smooth ($\Vc \in C^{1,2,1}$) and, by  construction, satisfies the HJB equation \eqref{eq:hjb}. As a result, the value function to Problem \eqref{eq:prob2} is indeed given by \eqref{eq:value}.
	The strategy $(\pi^*, L^*)$ given by \eqref{eq:op}  solves the supremum problem uniquely in the HJB equation \eqref{eq:hjb}, which is guaranteed by $\Vc_{xx}(t,x,y)= 2 A(t) < 0$.
	By Definition \ref{def:adm}, $(\pi^*, L^*)$  is admissible.
	Hence, we conclude that the strategy given by \eqref{eq:op} is optimal to Problem \eqref{eq:prob2}.
	The proof is now complete.
\end{proof}

\bibliographystyle{apalike}
\bibliography{MV-Investment-Risk-Final}

\end{document}